\documentclass[12pt,preprint]{emulateapj}
  
\usepackage{apjfonts}
\usepackage{amsmath}
\usepackage{epsf}
\usepackage{color}
\shorttitle{CANDELS: Contribution of Observed Galaxies to Reionization}
\shortauthors{Finkelstein et al.}

\newcommand{\sol}{$_{\odot}$}
\newcommand{\lya}{Ly$\alpha$}

\def\arcs{\hbox{$^{\prime\prime}$}}

\begin{document}
\slugcomment{Submitted to the Astrophysical Journal}
\title{CANDELS: The Contribution of the Observed Galaxy Population to Cosmic Reionization}

\author{Steven   L.    Finkelstein\altaffilmark{1,a}, Casey Papovich\altaffilmark{2}, Russell E. Ryan Jr.\altaffilmark{3},  Andreas H. Pawlik\altaffilmark{1}, Mark Dickinson\altaffilmark{4}, Henry C. Ferguson\altaffilmark{3}, Kristian Finlator\altaffilmark{5,a}, Anton M. Koekemoer\altaffilmark{3}, Mauro Giavalisco\altaffilmark{6}, Asantha Cooray\altaffilmark{7}, James S. Dunlop\altaffilmark{8}, Sandy M. Faber\altaffilmark{9}, Norman A. Grogin\altaffilmark{3}, Dale D. Kocevski\altaffilmark{9}, \& Jeffrey A. Newman\altaffilmark{10}}

\altaffiltext{1}{Department of Astronomy, The University of Texas at Austin, Austin, TX 78712}
\altaffiltext{2}{George P. and Cynthia Woods Mitchell Institute for Fundamental Physics and Astronomy, Department of Physics and Astronomy, Texas A\&M University, College Station, TX 77843}
\altaffiltext{3}{Space Telescope Science Institute, Baltimore, MD 21218}
\altaffiltext{4}{National Optical Astronomy Observatory, Tucson, AZ 85719}
\altaffiltext{5}{Physics Department, University of California, Santa Barbara, CA 93106}
\altaffiltext{6}{Department of Astronomy, University of Massachusetts, Amherst, MA 01003}
\altaffiltext{7}{Department of Physics and Astronomy, University of California, Irvine, CA 92697}
\altaffiltext{8}{Institute for Astronomy, University of Edinburgh, Royal Observatory, Edinburgh, UK}
\altaffiltext{9}{University of California Observatories/Lick Observatory, University of California, Santa Cruz, CA, 95064}
\altaffiltext{10}{Department of Physics and Astronomy and Pitt-PACC, University of Pittsburgh, Pittsburgh, PA 15260}
\altaffiltext{a}{Hubble Fellow, stevenf@astro.as.utexas.edu}

\begin{abstract}
We present measurements of the specific ultraviolet luminosity density
from a sample of 483 galaxies at $6 \lesssim z \lesssim 8$.  These
galaxies were selected from new deep near-infrared {\it Hubble Space
  Telescope} imaging from the Cosmic Assembly Near-infrared Deep
Extragalactic Legacy Survey, Hubble UltraDeep Field 2009 and WFC3
Early Release Science programs.  
In contrast to the majority of previous analyses, which assume that the
distribution of galaxy ultraviolet (UV) luminosities follows a
Schechter distribution, and that the distribution continues to
luminosities far below our observable limit, we investigate the
contribution to reionization from galaxies which we can observe, free
from these assumptions.  Due to our larger survey volume, wider wavelength coverage, 
and updated assumptions about the clumping of gas in the intergalactic
medium (IGM), we find that the observable population of galaxies can
sustain a fully reionized IGM at $z =$ 6, if the average ionizing
photon escape fraction (f$_{esc}$) is $\sim$30\%.  
A number of previous studies have measured UV luminosity densities at these
redshifts that vary by a factor of 5, with many concluding that galaxies
could not complete reionization by $z =$ 6 unless a large population
of galaxies fainter than the detection limit were invoked, or
extremely high values of f$_{esc}$ were present.
The observed UV luminosity density from our observed galaxy samples at $z =$ 7 and 8 is not sufficient to
maintain a fully reionized IGM unless f$_{esc}$ $>$ 50\%.
We examine the contribution from galaxies in different
luminosity ranges, and find that the sub-$L^{\ast}$  galaxies we detect
are stronger contributors to the ionizing photon budget than the $L >$
$L^{\ast}$  population, unless f$_{esc}$ is luminosity
dependent.  Combining our observations with constraints on the emission
rate of ionizing photons from Ly$\alpha$ forest observations at $z =$ 6, we find
that we can constrain f$_{esc} <$ 34\% (2$\sigma$) if the observed galaxies are the only
contributors to reionization, or $<$ 13\% (2$\sigma$) if the luminosity function
extends to a limiting magnitude of M$_{UV} = -$13.  These escape
fractions are sufficient to complete reionization by $z =$ 6.  
Current constraints on the high-redshift galaxy population imply
 that the volume ionized fraction of the IGM, while consistent with unity at $z \leq$ 6, appears
to drop at redshifts not much higher than 7, consistent with a number of
complementary reionization probes.
If faint galaxies dominated the ionizing photon budget at
$z =$ 6--7, future extremely deep observations with the {\it James
  Webb Space Telescope} will probe deep enough to see them, providing
an indirect constraint on the global ionizing photon escape fraction.
\end{abstract}


\section{Introduction}

The reionization of the intergalactic medium (IGM) is the last major
phase transition in the Universe, and as such has been a major topic
of recent study.  Over the past decade, a number of
lines of evidence have yielded improved constraints on the duration of
reionization, the completion redshift, and the primary source of
ionizing photons.

Observations of the Cosmic Microwave Background (CMB) measure the
Thomson scattering optical depth due to electrons in the path.  The
updated 7 year {\it Wilkinson Microwave Anisotropy Probe} results
measure $\tau =$ 0.088 $\pm$ 0.014, which corresponds to a redshift of
instantaneous reionization of $z =$ 10.6 $\pm$ 1.2 \citep{komatsu11}.
However, reionization is likely to be a much more extended event,
especially if faint galaxies play a dominant role.  Measurements of
the kinetic Sunyaev-Zeldovich (kSZ) effect with the South Pole Telescope
(SPT) from \citet{zahn11} have recently placed a limit on the duration of reionization, of $\Delta z
<$ 7.9.  When combined with the {\it Wilkinson Microwave Anisotropy
  Probe} ({\it WMAP}) constrains on instantaneous
reionization, they conclude at 95\% confidence that reionization was complete
at $z >$ 5.8.  Additionally, observations of the \lya\ forest in high-redshift
quasars also imply an end to reionization at $z \sim$ 6, as the
measurements of the near-zones around $z =$ 6 quasars yield results
consistent with a negligible neutral fraction \citep[e.g.,][]{fan06},
though more $z >$ 6 quasar sight-lines may be needed to conclusively verify this result
\citep{mesinger10} \citep[see also][]{becker07}.

Nonetheless, some sources must be responsible for reionizing the IGM between 6
$\lesssim$ z $\lesssim$ 14.  Quasars were a natural choice, as they
are extremely bright, and the bulk of their ionizing photons can
escape.  However, the quasar luminosity function peaks at $z \sim$ 2,
and falls off rapidly toward higher redshift
\citep[e.g.,][]{hopkins07}.  Additionally, observations of the X-ray
background rule out a dominant contribution to the reionizing photon
budget from quasars \citep{dijkstra04b}.  Population III stars, due to
their predicted high masses \citep[e.g.,][]{bromm04,glover05}, were
very efficient emitters of ionizing photons \citep{tumlinson00,
  bromm01, schaerer02}. Their overall contribution to reionization
may, however, have been limited \citep[e.g.,][]{greif06}. This conclusion has been
strengthened by recent simulation results which have corrected
the mass scale of the first stars down to less extreme values
\citep{clark11,greif11,greif12}.

The most likely source for the bulk of ionizing photons is thus from star-forming
galaxies themselves, where ionizing photons are created from the massive stars
present during ongoing star formation.  This has been well-studied
observationally, but it was only in the last decade when large samples of
galaxies close to the reionization epoch were compiled.  Following
wide and deep surveys with the Advanced Camera for Surveys (ACS)
onboard the {\it Hubble Space Telescope} ({\it HST}), large samples of
$z \sim$ 6 galaxies were compiled
\citep[e.g.,][]{giavalisco04,bunker04}.  While \citet{giavalisco04}
found that the specific UV luminosity density (and thus the
star-formation rate density) was roughly constant from $z =$ 4 to 6,
\citet{bunker04} found, via deeper observations, that this quantity
appeared to decline towards higher redshift.  Using the commonly accepted (at the time) large
value for the clumping factor of the IGM, they concluded that galaxies
were unable to account for the necessary ionizing photons to reionize
the universe.  \citet{yan04} pointed out that if the faint-end slope
of the galaxy luminosity function was steep, it could be that $L <
0.1L^{\ast}$  galaxies dominated the ionizing photon budget, and thus
allowed galaxies to complete reionization by $z =$ 6.  Regardless of
the interpretation, these
galaxy samples contained only $\sim$ 50-100 galaxies, each of which
was only detected in one band, given the lack of deep {\it HST}
near-infrared observations at the time.

The advent of the Wide Field Camera 3 (WFC3) on {\it HST} has opened the door to
the $z \geq$ 6 universe, allowing the first investigation into whether
reionization completed at $z =$ 7 or earlier.  This is combined with
recent results which have vastly changed the
expected IGM clumping factor to much lower values.
\citet{finkelstein10a} investigated the combined UV luminosities of
their observed $z =$ 7 galaxies, and found that while they could
come close to reionizing the IGM, either fainter galaxies or high
($\geq$ 50\%) escape fractions were necessary to complete reionization
by $z =$ 7 \citep[see also][]{bunker10}

A number of studies have used measured luminosity functions of $z
\sim$ 7 and 8 galaxies to infer their contribution to reionization, by
integrating them down to an assumed magnitude limit.  The majority of these studies found that, once galaxies fainter than
the detection threshold are accounted for, galaxies could reionize the
universe at such early times \citep[e.g.,][]{oesch10,mclure10,lorenzoni11,bouwens11d}
\citep[see also][]{grazian11}.  This is however reliant on
a number of assumptions.  First, the luminosity function results are
susceptible to the assumption of a Schechter function
parameterization.  At $z \sim$ 5--6, where the samples are larger,
there is no strong evidence for deviation from this function
\citep[e.g.,][]{bouwens07,mclure09}.
However, as we push closer to the Big Bang, galaxies are changing
rapidly, thus at some point we may encounter an epoch where the
Schechter function is no longer an accurate representation.  This
could be due to a variety of effects, with one being a lack of
active galactic nuclei (AGN) feedback if AGNs are not yet present in
the centers of all galaxies at very high redshifts, which is plausible
depending on the speed with which supermassive black holes form in the
early universe.  

Additionally, while recent evidence has indicated a very steep faint end slope at $z
\geq$ 7 \citep[e.g.,][]{bouwens11, oesch12, bradley12}, the
uncertainty on these measurements are large \citep[e.g., $\sigma
(\alpha) \approx$ 0.2 at $z =$ 8][]{bradley12}.  Finally, when integrating the luminosity function, one needs to choose a limiting
magnitude, as suppression from the UV background will result
in the gas in galaxies becoming heated below some circular velocity (i.e. mass) limit.
Galaxies above this limit are dense enough where collisional
excitation of H\,{\sc i} by electrons can overcome photoionization heating, and
have their gas cool and form stars.  However, we have no observational
evidence for this value, and theoretical results from the literature
yield values of $-$15 $< M_{lim} < -$10 \citep[e.g.,][]{finlator11b,munoz11,kulkarni11,choudhury08}.  With a steep
faint-end slope, a difference of this level can result in a difference
in the integrated luminosity density by more than a factor of two.

Here we measure the specific luminosity density of the observable galaxy
population  using the  largest sample  of  $6 <  z <  8$ galaxies  yet
compiled, which  allows us  to study the  contribution of  galaxies to
reionization  without   the  uncertainties  inherent   in  invoking  a
parametrized luminosity  function.  This allows us to assess whether a
significant contribution from galaxies below the observational limits
is necessary to complete reionization at a given redshift.  
We emphasize our results at $z =$ 6, where we have a large sample, and
each galaxy is detected in four individual imaging bands, yielding
much more robust results over the previous studies which had access
only to optical ACS data.

In \S 2 we describe the datasets
used  in this study,  as well  as our  photometry and  sample selection
methods.  We discuss  how  we measured  the rest-frame  UV
specific  luminosity  density,  and  corrected  it  for
incompleteness down to  our magnitude limit in \S 3, while in \S  4, we discuss the
implications  of our  results  on the  reionization  of the  IGM.
Throughout  this  paper  we   assume  a  concordance  cosmology,  with
H$_\mathrm{o}$  = 70  km  s$^{-1}$ Mpc$^{-1}$,  $\Omega_m$  = 0.3  and
$\Omega_\lambda$ = 0.7.  All magnitudes are reported in the AB system,
where  $m_\mathrm{AB}  = 31.4  -  2.5\log(f_\nu  / 1\,  \mathrm{nJy})$
\citep{oke83}.  In the remainder of the paper we make use of
luminosity functions from the literature over the redshift range of
our study.  We use the luminosity functions from \citet{bouwens07} for
$z =$ 4, 5 and 6; \citet{bouwens11} for $z =$ 7; \citet{bradley12} for
$z =$ 8 and \citet{bouwens11c} for $z =$ 10.  These represent the
most up-to-date luminosity functions for these redshifts at the time
of this writing.  The characteristic magnitude values from these
studies are: $M^{\ast}$ = $-$20.98 ($z =$ 4), $-$20.64 ($z =$ 5),
$-$20.24 ($z =$ 6), $-$20.14 ($z =$ 7), $-$20.26 ($z =$ 8) and $-$18.3 ($z =$ 10). 

\vspace{5mm}

\section{Observations}

In this Paper, we use a sample of high-redshift galaxies selected
from {\it HST} data in the GOODS-S region.  We review our methods
below, and note that they are similar to those used in \citet{finkelstein11c}, thus the
reader is referred there for further details.

\begin{deluxetable*}{cccccccccccc}
\tabletypesize{\small}
\tablecaption{Observations Summary}
\tablewidth{0pt}
\tablehead{
\colhead{Field} & \colhead{Area} & \colhead{B$_{435}$} & \colhead{V$_{606}$} & \colhead{i$_{775}$} & \colhead{I$_{814}$}  & \colhead{z$_{850}$} & \colhead{$Y_{098}$} & \colhead{$Y_{105}$} & \colhead{$J_{125}$} & \colhead{$H_{160}$}\\
\colhead{$ $} & \colhead{(arcmin$^{2}$)} & \colhead{(mag)} & \colhead{(mag)} & \colhead{(mag)} & \colhead{(mag)} & \colhead{(mag)} & \colhead{(mag)} & \colhead{(mag)} & \colhead{(mag)} & \colhead{(mag)}\\
}
\startdata
HUDF09 MAIN&5.0&29.5&29.9&29.7&---&29.0&---&29.2&29.5&29.5\\
HUDF09 PAR1&4.4&---&28.9&28.7&---&28.6&---&28.0&28.9&28.7\\
HUDF09 PAR2&4.6&---&29.0&28.7&---&28.4&---&28.7&29.1&28.9\\
GOODS-S DEEP&62.0&28.1&28.3&27.7&28.2&27.5&---&28.2&28.1&27.9\\
GOODS-S WIDE&34.1&28.1&28.3&27.7&27.8&27.5&---&27.4&27.6&27.3\\
GOODS-S ERS&40.5&28.1&28.3&27.7&27.7&27.5&27.6&---&28.0&27.7\\
\enddata
\tablecomments{The area is measured from the image area used to
  detect objects, which excludes noisy regions on the edges, regions
  without data from all available filters, as well
  as regions overlapping deeper data (i.e., the CANDELS Deep field
  also covers the HUDF, thus we have masked out this region in the
  CANDELS imaging for this analysis).   The
  remaining columns are 5$\sigma$ limiting magnitudes measured in
  a 0.4\arcs-diameter apertures on non-PSF matched images.  The F814W
  data from CANDELS was only used in the visual inspection of $z =$ 8
  candidate galaxies, primarily in the GOODS-S DEEP field as it is
  $\sim$0.5 mag deeper than the existing GOODS ACS data.}
\end{deluxetable*}

\subsection{Data}
We use WFC3 imaging data from three separate programs.  The first is
the HUDF09 program (PID 11563; PI Illingworth), which obtained deep
imaging over the Hubble UltraDeep Field (HUDF) as well as its two
flanking fields (which we will refer to as HUDF09-01 or PAR1 and HUDF09-02
or PAR2).  The second dataset comes from the WFC3 Early
Release Science Program \citep[ERS; PID 11359; PI O'Connell;][]{windhorst11}, which imaged
the northern $\sim$25\% of the GOODS-S field.  The third dataset comes
from the Cosmic Assembly Near-Infrared Deep Extragalactic Legacy
Survey \citep[CANDELS; PI Faber \& Ferguson;][]{grogin11,koekemoer11}, which obtained deep imaging of
the central $\sim$50\% of GOODS-S (hereafter referred to as CANDELS
DEEP), and less deep imaging of the southern $\sim$25\% of GOODS-S
(hereafter referred to as CANDELS WIDE; PID 12060, 12061 \& 12062).  We
use the completed CANDELS dataset in GOODS-S, which includes
10 epochs in the DEEP field, as well as both epochs of WIDE imaging.  All three of these surveys obtained
imaging in the F125W and F160W filters (hereafter J$_{125}$ and
H$_{160}$).  The CANDELS and HUDF09 surveys also obtained imaging in
the F105W filter, while the ERS used the narrower F098M filter
(hereafter Y$_{105}$ and Y$_{098}$, respectively).  These fields and
their associated depths are summarized in Table 1.

\subsection{Photometry and Sample Selection}

We performed photometry on all three datasets using the Source
Extractor software \citep{bertin96}.  We chose a single set of
parameters that we applied to all images, which we tested to maximize
detection of real sources while minimizing contamination
(the key parameters are DETECT\_MINAREA$=$7 and DETECT\_THRESH$=$0.6).
We did this in two image mode,
using a weighted sum of the J$_{125}$ and H$_{160}$ images as the
detection image, and the three WFC3 bands, as well as the archival
GOODS-S ACS F435W, F606W, F775W and F850LP (hereafter B$_{435}$, V$_{606}$,
i$_{775}$ and z$_{850}$, respectively) bands as the measurement
images.  We also measured photometry on very deep F814W ACS imaging,
obtained in parallel during the CANDELS observations.  Recently
obtained ACS imaging suffers from poor charge transfer efficiency; this can be corrected
for detectable sources \citep{anderson10}, but at this point it is
unclear whether robust upper limits can be obtained from these data.
We thus use the F814W imaging (which has had this correction applied) only as a veto band when inspecting $z
=$ 8 galaxy candidates, as at these redshifts there should be no
detectable flux in this band.  All images were matched to the
point-spread function (PSF) of the H$_{160}$ image prior to the
photometry.

We selected our sample of high-redshift galaxies using the photometric
redshift fitting software EAZY \citep{brammer08}.  Rather than using
the best-fit photometric redshift, we used the full probability
distribution function (PDF) to select our galaxies.  We do this by
compiling samples in three redshift bins, $z =$ 6, 7 and 8.  To make
it into a given bin, an object must have a $\geq$ 3.5$\sigma$
detection in {\it both} the J$_{125}$ and H$_{160}$ bands, it must
have $\geq$ 70\% of its integrated PDF in the primary redshift solution, and it
must have $\geq$25\% of its integrated PDF within $z_{sample} \pm$ 0.8 (i.e., for
$z =$ 6, $\geq$25\% of the integrated PDF must be at 5.2 $\leq$ z $\leq$ 6.8).  Finally,
objects which satisfy this criterion for multiple samples are placed
in the sample which contains a larger fraction of the PDF (for this
purpose, we also integrated the PDF for hypothetical $z =$ 5 and $z =$
9 samples, to prevent those objects from being placed in our $z =$ 6
and $z =$ 8 samples, respectively).  These methods are similar to that used in
\citet{finkelstein11c}, and their Figure 1 illustrates this
selection technique.

This selection was performed separately for each imaging dataset.
All selected candidates were visually inspected, to screen against
false detections such as diffraction spikes or oversplit regions on
the edges of bright galaxies.  Additionally, we screened against
stellar contaminants, first by investigating whether a given source is
resolved, which we did by comparing the measured full-width at
half-maximum (FWHM) of the sources to that of stars in the image (where this
measurement was performed by Source Extractor).  For sources that had
FWHMs close to that of stars in the image, we compared their colors to
measured colors of late-types stars and brown dwarfs, integrating
observed NASA Infrared Telescope Facility (IRTF) SpeX prism spectra through the ACS and WFC3
bandpasses\footnote[1]{This research has benefitted from the SpeX
  Prism Spectral Libraries, maintained by Adam Burgasser at
  http://pono.ucsd.edu/$\sim$adam/browndwarfs/spexprism.}.  We found
15 likely M-dwarfs in our $z =$ 6 sample, and 4 likely brown dwarfs in
our $z =$ 7 sample.  After clearing these contaminants from the
sample, we had a total of 302, 136 and 45 galaxy candidates at $z =$ 6, 7
and 8, respectively.  These samples are summarized in Figure 1, which highlights the
importance of a wide area at $z =$ 6, and very deep imaging at $z =$ 8.

\begin{figure}[!t]
\epsscale{1.2}
\plotone{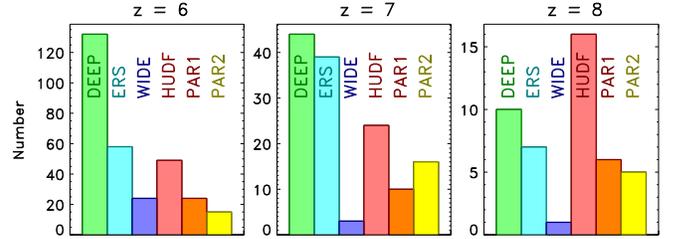}
\caption{The number of objects selected in each of our six fields for
$z =$ 6, 7 and 8, from left-to-right.  At $z =$ 6 -- 7, the
CANDELS Deep and ERS fields dominate the sample.  However, at $z =$ 8,
due to the decrease in the characteristic magnitude $M^{\ast}$ and the
increased luminosity distance, the deeper HUDF09 fields dominate the
sample.  The CANDELS Wide field contributes a non-negligible number of
galaxies at $z =$ 6, but is not deep enough to contribute
significantly at higher redshifts.}
\end{figure}  

\section{Analysis}

\begin{figure*}[!t]
\epsscale{0.85}
\plotone{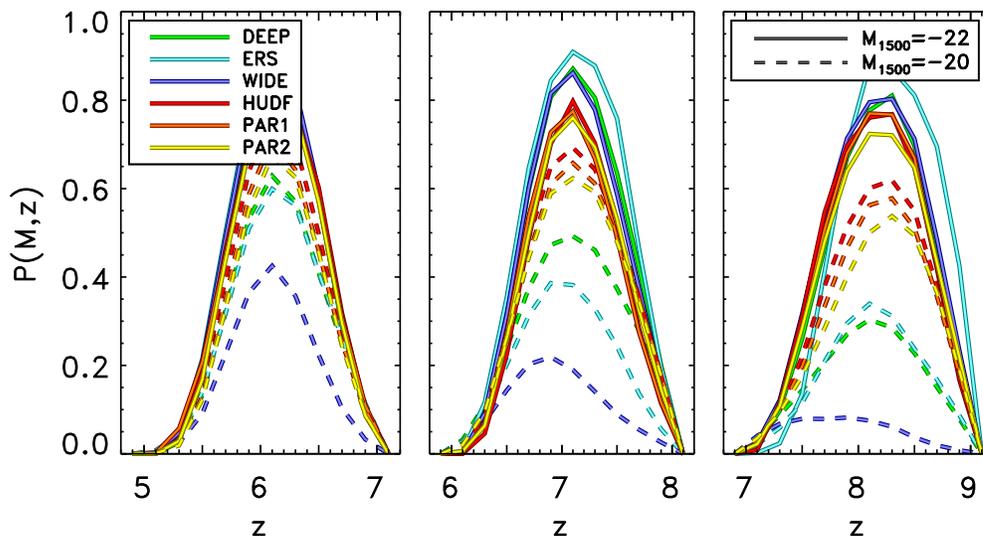}
\vspace{-2mm}
\caption{A summary of our completeness simulations.  Each panel
  represents one of our redshift bins of $z =$ 6, 7 and 8 from
  left-to-right.  Each panel contains the probability of recovering a
  given galaxy (as a function of redshift and $M_{1500}$) for two
  magnitude values; $M_{1500}$ = $-$22 (solid line), and $M_{1500}$ =
  $-$20 (dashed line).  At $z =$ 6, 7 and 8 $M_{1500}$ =
  $-$22 corresponds to apparent UV magnitudes of 24.7, 24.9 and 25.1, and $M_{1500}$ = $-$20 corresponds to
  26.7, 26.9 and 27.1, respectively.  While all fields can efficiently detect bright
galaxies at any redshift, only the two HUDF09 fields can efficiently
detect fainter galaxies at higher redshift, which is consistent with
our recovered distribution shown in Figure 1.  Although the HUDF09
fields are more efficient than the CANDELS Deep and ERS fields for
fainter galaxies at $z
=$ 6, the much larger area probed by the latter fields results in
their dominating the sample at that redshift.  We note that although they
are of comparable area, the volume probed by the WIDE field is much
less than the ERS.  At fainter magnitudes, this is due to the
increased depth in the ERS.  However, even at bright magnitudes, the
ERS is more complete at $z =$ 7 and 8, as the F098M filter used in the ERS does not overlap with
F125W (while F105W, used in the remaining fields, does), allowing more
precise photometric redshifts.}
\end{figure*} 

\subsection{Measuring the Rest-Frame UV Light}

In order to study the contribution of the galaxies in our sample to
the reionization of the universe, we need to measure their rest-frame
UV absolute magnitudes.  We have chosen to do this in the same manner
as in \citet{finkelstein11c}, which we briefly summarize here.  We
first compute a grid of synthetic stellar population models over a
range of redshifts, population ages, stellar metallicities,
star-formation histories and dust attenuations, using the updated
models of \citet[][hereafter BC03]{bruzual03}.  We then found the best fit model for a
given galaxy in our sample via $\chi^2$ minimization, and then integrated this best-fit model
spectrum through a 100 \AA-wide square bandpass centered at rest-frame 1500 \AA\ to
compute the flux at that wavelength.  This was then converted to the
absolute magnitude at 1500 \AA\ ($M_{1500}$) using the distance
modulus to the best-fit photometric redshift.  The uncertainty on
$M_{1500}$ was derived via a series of 100 Monte Carlo simulations,
where in each simulation each galaxy's fluxes were altered by a
Gaussian random distribution with a mean of zero and a standard
deviation equal to the photometric error for the band in question.
In each simulation, a new redshift was chosen based on the photometric
redshift PDF for a given object; in this way the uncertainties on all
derived parameters include the inherent uncertainty in the redshift.
The 68\% confidence range of $M_{1500}$ was then derived as the central 68\%
spread of $M_{1500}$ values from the simulations for each object.

\subsection{Incompleteness Correction}
Prior to computing the specific luminosity density, we first need to
correct for incompleteness down to our limiting magnitude.  We have
done this by inserting mock galaxies into our images, running separate
simulations for each redshift bin and each dataset (for a total of 18
sets of simulations).  In each simulation, we chose a uniform random
distribution of redshifts ranging from $z_{sample}$ $\pm$ 1, as well
as a range of H$_{160}$ magnitudes modeled after the number counts in
a given field, but continuing the trend to fainter magnitudes.  We
then chose a stellar metallicity, population age, and color excess
E(B-V) from distributions which were tuned such that the recovered galaxies in a given
simulation had a similar rest-frame UV spectral slope distribution as
those of the recovered galaxy sample (ensuring that we're not
correcting for, e.g., a population of red galaxies that do not exist
in reality).  A stellar population model was then created with these parameters, and
integrated through the bandpasses of interest to derive galaxy
magnitudes in each band.

The image of each mock galaxy was created with the GALFIT software
package \citep{peng02}.  The position angle was drawn from a uniform
random distribution, while the S\'{e}rsic index $n$ was drawn from a
log-normal distribution, such that the mock sample was predominantly
disk galaxies, with fewer bulge-dominated galaxies.  Both the
half-light radius and the axial ratio were tuned such that the
recovered distributions matched the distributions from our observed
galaxy sample (to prevent against, i.e., correcting for large galaxies which do not exist).

Samples of 100-200 mock galaxies were placed in sub-regions of the
images in each field.  The images were then treated in the same manner
as in \S 2.2, in that we derived photometry with Source Extractor, and
photometric redshifts were measured with EAZY.  Samples were selected
at each redshift using the same criteria as that used on our galaxy
samples.  This process was repeated until $\sim$ 10$^{5}$ galaxies were
input and measured from the images (i.e., 500-1000 iterations).  Typical numbers of recovered
galaxies were $\sim$ 2 -- 3 $\times$ 10$^{4}$ (depending on the
field and redshift), where a galaxy is
denoted as recovered if it is detected in the photometry and selected
as a candidate in the redshift range of interest.

For each field and at each redshift, we then computed the probability
that a given galaxy was recovered in our sample as a function of the
input redshift and $M_{1500}$ (where here $M_{1500}$ was measured from
the input spectrum to the simulations), defined as
$\mathcal{P}(M,z)$.  These values are shown at a few representative magnitudes in Figure 2.

\subsection{Evolution of the Rest-Frame UV Specific Luminosity Density}

Using the value of $M_{1500}$ for each galaxy, as well as
$\mathcal{P}(M,z)$ for each field and redshift, we computed the
rest-frame UV specific luminosity density ($\rho_{UV}$), in units of
erg s$^{-1}$ Hz$^{-1}$ Mpc$^{-3}$ (specific refers to per
unit frequency, while density refers to per unit volume).  We converted the
absolute magnitude $M_{1500}$ of each galaxy to the 1500
\AA\ specific luminosity (i.e., erg s$^{-1}$ Hz$^{-1}$) using the
best-fit photometric redshift (while noting that the uncertainty in
$M_{1500}$ includes the uncertainty in the redshift; see \S 3.1). 

We computed the effective volume, which is the volume over which we are
selecting galaxies given our imaging depths and selection criteria,
defined as
\begin{equation}
V_{eff}(M_{1500})_{field} = A_{field} \int \frac{dV}{dz}~\mathcal{P}(M_{1500},z)~dz
\end{equation}
where A$_{field}$ is the area probed by a given field, which is given
in Table 1, and $\frac{dV}{dz}$ is the co-moving volume per square
arcminute per redshift slice $dz$ at the redshift $z$.  The value of
$\rho_{UV}$ is thus the sum of the specific luminosities divided by
the effective volume for a given field and redshift, in each magnitude bin.
We note that in the following, when we refer to measurements of our observed galaxy
sample, or the observable galaxy sample at a given redshift, we refer
to these incompleteness-corrected measurements.

We computed a total value of $\rho_{UV}$ for each
redshift, by first taking the sum of all specific luminosities in a given redshift
sample in a given magnitude bin, and then dividing it by the sum of
the effective volumes for each field in that magnitude bin.  We
corrected down to a limiting M$_{1500} = -$18, which corresponds to
0.13, 0.14 and 0.19 $\times$ $L^{\ast}$  at $z =$ 6, 7 and 8,
respectively (as well as 0.06 $\times$ L$^{\ast}_{z=3}$; \citet{steidel99}).
The total value of $\rho_{UV}$ was then the sum of these values over all
magnitude bins.  These values are shown as the large cyan circles in
Figure 3, and are tabulated in Table 2.

The uncertainties quoted on $\rho_{UV}$ are comprised of three
components: Poisson noise, photometric scatter, and uncertainty in the
effective volumes.  These were included simultaneously in a series of
10$^{4}$ bootstrap Monte Carlo simulations.  In each simulation, we
accounted for Poisson noise by selecting $N^{\prime} = N + A \times \sqrt{N}$ galaxies in each
redshift bin through random sampling with replacement, where N is the
number of galaxies in the bin, and A is a random number drawn from a
normal distribution with a mean of zero and a standard deviation of
unity.  We then account for the photometric noise by taking each of
the galaxies in the modified sample, and obtaining a new estimate,
$M_{1500}^{\prime}$, for $M_{1500}$, where $M_{1500}^{\prime}$ is the
original value $M_{1500}$ modified by a Gaussian with $\sigma$ equal
to the photometric error (performed in
flux-space).  Finally, we accounted for the uncertainty in the
effective volume simulations by computing the uncertainty in
$\mathcal{P}(M_{1500},z)$  via a separate set of 
simulations, randomly selecting N recovered simulated galaxies, where
$N$ is 50\% of the total number of recovered galaxies.  This was done
10$^{3}$ times, generating 10$^{3}$ $\mathcal{P}(M_{1500},z)$ arrays.  The uncertainty in $\mathcal{P}(M_{1500},z)$
was derived as the standard deviation of the 10$^{3}$
probability values at each magnitude and redshift.  These
uncertainties were then folded in to our uncertainty on $\rho_{UV}$
during the Monte Carlo simulations by modifying each value of the 
$\mathcal{P}(M_{1500},z)$ array by a Gaussian with $\sigma$ equal
to the uncertainty in $\mathcal{P}(M_{1500},z)$.
Thus, at each redshift we obtained 10$^{4}$ values for $\rho_{UV}$ including the effects of
sample size, photometric scatter, and uncertainty in the effective
volumes.  The uncertainty we quote is
the central 68\% range of these values.

We list the total effective volumes probed by our study at each
redshift (at two representative magnitudes) in Table 2.  We note that
these volumes are subject to our assumptions of the galaxy properties
in our incompleteness simulations.  As discussed in the previous
subsection, the key parameters in recovering a galaxy are the color and
size.  While we have chosen what we believe are appropriate
distributions of both quantities (see \S 3.2 for discussion), it is
however interesting to examine how the effective volume would change if
the assumptions were different.  We have thus re-run our simulations
at $z =$ 6 in each field, using a modified color and size
distribution (in comparison to our default distributions, we broadened the distribution of dust attenuation, and
changed the size distribution to peak at smaller sizes, but with a
higher large-radius tail).  We found that the derived effective
volumes from this modified simulation were $\sim$ 10\% smaller than our
nominal simulations in \S 3.2.  Thus, the derived UV luminosity
densities would be $\sim$ 10\% higher if these volumes were more
physically correct.  However, as discussed above, we believe our
assumed physical property distributions are physically robust, so our
effective volumes should be representative of the true volumes probed
by our survey.  As this test shows, if for whatever reason this is not
the case, it creates a systematic uncertainty on $\rho_{UV}$ on the
order of 10\% for modest changes in the input distributions.

\begin{deluxetable*}{cccccccc}
\tabletypesize{\small}
\tablecaption{Rest-Frame UV Specific Luminosity Density}
\tablewidth{0pt}
\tablehead{
\colhead{z} & \colhead{N$_{Galaxies}$} & \colhead{$\rho_{UV}$} & \colhead{$\rho_{UV}$} & \colhead{$\rho_{UV}$} & \colhead{$\rho_{UV}$} & \colhead{$V_{eff}$ ($M_{1500} = -$22)} & \colhead{$V_{eff}$ ($M_{1500} = -$20)}\\
\colhead{$ $} & \colhead{$ $} & \colhead{M$_{1500} < -$18} & \colhead{L $>$ 0.2L$^{\ast}$} & \colhead{L $>$ 0.5L$^{\ast}$} & \colhead{L $>$ L$^{\ast}$} & \colhead{(Mpc$^{-3}$)} & \colhead{(Mpc$^{-3}$)} \\
}
\startdata
6&302&0.79 $\pm$ 0.06&0.67 $\pm$ 0.05&0.44 $\pm$ 0.04&0.25 $\pm$ 0.03&3.07 $\pm$ 0.06 $\times$ 10$^5$&2.28 $\pm$ 0.03  $\times$ 10$^5$\\
7&136&0.44 $\pm$ 0.04&0.43 $\pm$ 0.04&0.27 $\pm$ 0.03&0.17 $\pm$ 0.02&3.04 $\pm$ 0.04 $\times$ 10$^5$&2.23 $\pm$ 0.02 $\times$ 10$^5$\\
8&45&0.27 $\pm$ 0.06&0.27 $\pm$ 0.06&0.14 $\pm$ 0.03&0.07 $\pm$ 0.02&2.96 $\pm$ 0.05 $\times$ 10$^5$&1.94 $\pm$ 0.02 $\times$ 10$^5$\\
\enddata
\tablecomments{All specific luminosity density values are in the units
of 10$^{26}$ erg s$^{-1}$ Hz$^{-1}$ Mpc$^{-3}$.  The effective volumes
are quoted in comoving units, and were derived from the completeness simulations
discussed in \S 3.2.  The uncertainties on the volumes come from
bootstrap simulations, succesively selecting 50\% of the recovered
simulated galaxies to assess the variation in $\mathcal{P}(M,z)$.}
\end{deluxetable*}

\subsection{Critical Specific Luminosity Density to Complete Reionization}
As shown in Figure 3, our results show a decrease with redshift in the specific
luminosity density for the observable galaxy population down to
$M_{1500} < -$18.  In order to understand
the implications of this trend on the reionization of the IGM, we need
to connect our observable, $\rho_{UV}$, to the necessary number of
ionizing photons to sustain a fully reionized IGM.  We do this in the same manner as
in \citet{finkelstein10a}, where we followed \citet{madau99} to
derive the required 1500 \AA\ luminosity density to sustain
reionization:
\begin{equation}
  \begin{split}
    \rho_\mathrm{UV} & = 1.24 \times 10^{25}\,\, \epsilon_{53}\,^{-1} \left(
      \frac{1+z}{8} \right)^3  \left( \frac{\Omega_b h^2_{70}}{0.0461}
    \right)^2\\ 
  &  \times x_{HII}\,\, \frac{C}{f_\mathrm{esc}}\,\,\, \mathrm{erg}\,\,\mathrm{s}^{-1}\,\,\mathrm{Hz}^{-1}\,\,\mathrm{Mpc}^{-3},
  \end{split}
\end{equation}
where $\Omega_b$ is the cosmic baryon density and $h_{70}$ is the
Hubble parameter in units of $h=0.7$, and the constant 0.0461 is our
assumed value of $\Omega_b$ ($\times$ $h_{70}^2$) from {\it Wilkinson
  Microwave Anisotropy Probe} Year 7 data \citep{komatsu11}.  The
clumping factor of ionized hydrogen in the IGM is denoted by C, and ties in
the average rate of recombinations in the IGM.  The
average escape fraction of ionizing photons from galaxies is denoted
by f$_{esc}$.  The variable $\epsilon_{53}$ is the number of Lyman continuum photons per
unit of forming stellar mass in units of 10$^{53}$ photons s$^{-1}$
(M\sol~yr$^{-1}$)$^{-1}$.  To determine this value, we used the
updated BC03 models with a Salpeter IMF (with stellar masses from 0.1
-- 100 M\sol) and a constant star-formation history.  We
examined values of the stellar metallicity from 0.02 -- 1.0 Z\sol\,
finding that $\epsilon_{53}$ ranges from 0.9 to 1.4\footnote[2]{These
  values assumed an age of 50 Myr.  For a constant star-formation
  history, $\epsilon_{53}$ is invariant for ages larger than
  10 Myr.}.  The volume ionized fraction is denoted by $x_{HII}$.  We
start by assuming this is equal to unity, and investigate changes in
\S 4.6.  The gray curves in Figure 3 denote the critical specific UV luminosity density from
this equation required to sustain a fully reionized IGM at a given redshift,
where the width of the curve denotes the range of $\epsilon_{53}$
values for differing metallicities.  Should the IMF be more top heavy than Salpeter,
or the metallicities in these galaxies be $\ll$ 0.01 Z\sol, it would increase the number
ionizing photons per observed UV photon, making it easier for a given
population to reionize the IGM \citep[e.g.,][]{chary08}.  We note that currently there
is no strong evidence to support such very low metallicities of these
galaxies \citep[e.g.,][]{bouwens10b, finkelstein10a, 
  finkelstein11c, wilkins11, finlator11b, wise12}.  

We reiterate that, under the assumption of $x_{HII} =$ 1, these curves
show the luminosity density required to maintain a fully ionized IGM.  Thus, if at a given
redshift our galaxies do not reach this threshold, it may not mean
that sources of ionizing photons are missing, it may just indicate the
epoch when the neutral fraction is no longer negligible.  As the
ionized fraction falls, one can calculate the luminosity density
necessary to maintain a given volume ionized fraction by lowering the
value of $x_{HII}$.  Previous observations constrain $x_{HII}$ to equal
unity at $z <$ 6 \citep[e.g.,][]{fan06}, and it is likely negligible by $z >$ 15 (c.f. \S 4.6).  As we are just
pushing back into the reionization epoch, we assume this parameter equals unity, but we note
that by $z \sim$ 8 it may be substantially reduced \citep[possibly to $\sim$0.5 -- 0.7,
e.g.,][]{bolton07,pritchard10,bolton11,mortlock11,pentericci11}.  The
primary effect of this is that a lower ionized volume fraction makes it easier for galaxies to
keep the ionized portion of the IGM in an ionized state.  The reader can take this into account when
investigating the various gray curves on Figure 3.

\section{Discussion}

\subsection{Assumptions}
The critical curves for sustaining a fully reionized IGM are shown as the gray
bands in Figure 3, for different assumptions of the ratio between the
clumping factor and escape fraction of ionizing photons, C/f$_{esc}$.
  The clumping factor of ionized hydrogen in the IGM is derived
  theoretically, and up until recently had been thought to be very
  large, C $\sim$ 30 \citep[e.g.,][]{gnedin97}.  This high value
  necessitates more ionizing photons to reionize a given volume,
  leading a number of studies to conclude that galaxies could not reionize the
universe \citep[e.g.,][]{bunker04}.  However, recent higher-resolution
simulations yield much lower values, of C $\sim$ 1 -- 5, primarily due
to a better understanding of the interface between the IGM and the
circum-galactic medium, but also because of an improved understanding of the feedback from
reionization \citep[e.g.,][]{miralda00,pawlik09,shull11} \citep[see
also][]{wise05,iliev06,raicevic11} and improvements in our
understanding of self-shielding \citep[e.g.,][]{mcquinn11}.  Recent
simulations by Finlator et al.\ (2012, in prep) show that when you only
consider gas that is sufficiently under-dense that its ionization state is determined by the
global UV background rather than by the next nearest galaxy (the impact of circum-galactic
gas being more appropriately attributed to f$_{esc}$),
C $\sim$ 3 at $z =$ 6, dropping to $\sim$ 2 by $z =$ 8.

The escape fraction of ionizing photons from galaxies is also poorly understood, and is one of the
primary goals of a number of observational studies.  This is
impossible to directly observe at the redshifts
of our study, as any escaping ionizing photons will be absorbed by
lingering neutral gas in the IGM.  Direct detection is possible at $z
<$ 4, but the results have been mixed.  \citet{siana10} searched for ionizing
photons with UV imaging of $z \sim$ 1 star-forming galaxies, finding
only non-detections, resulting in an upper limit on the relative
escape fraction of ionizing photons to UV photons of f$_{esc,rel} <$
2\% \citep[see also][]{siana07}.  At $z \sim$ 3, \citet{shapley06}
found that 2/14 LBGs have relative escape fractions of $\sim$
50--100\%, with the remainder having f$_{esc,rel} <$ 25\%.  A number
of other studies have detected escaping ionizing photons at $z \sim$
3--4 \citep[e.g.,][though we note some of these studies observed in the same
field (SSA22), thus they may have sources in common]{steidel01,iwata09,vanzella10,nestor11}, though space-based
imaging may be necessary to exclude contamination from intervening sources  \citep[e.g.,][]{vanzella12}.

Thus, it appears is if the average escape fraction of ionizing photons may be
increasing with increasing redshift (or, alternatively, the fraction of
galaxies exhibiting f$_{esc}$ significantly larger than zero rises
with redshift).  Future results from UV observations from CANDELS, as well as in the HUDF 
(PI Teplitz) will shed more light on this issue, but there are a few
observational clues to support this interpretation.  
A recent study by \citet{kuhlen12} have combined all
observational evidence (luminosity functions, {\it WMAP}, kSZ and \lya\
forest measurements) to create a concordance model of reionization,
which implies an increasing escape fraction at higher redshift
with f$_{esc}$ higher by perhaps as much as 10$\times$ at $z >$ 4.
Additionally, the fraction of galaxies exhibiting \lya\ emission (with a rest-frame \lya\
equivalent width $>$ 20 \AA) rises with redshift,
from $\sim$25\% at $z \sim$ 3 \citep[e.g.,][]{shapley03} to $\gtrsim$
60\% at $z \sim$ 6 \citep[e.g.,][]{stark10,stark11}.  Whatever
physical mechanism allows a larger fraction of \lya\ photons to escape
will likely also allow a larger fraction of ionizing photons to
escape (to a point; if all ionizing photons escape, no \lya\ will be
produced).  One possible scenario is that the covering fraction of
neutral gas decreases with increasing redshift.  Observational
evidence for this has been found, as \citet{jones11} measured that
the low ionization absorption lines in a stacked rest-frame UV
spectrum of $z =$ 4 galaxies were weaker than at $z =$ 3, implying a
lower covering fraction of neutral gas at higher redshift.
Additionally, they stacked galaxies with and without \lya\ emission,
and found weaker absorption features in the stack of galaxies with
\lya\ emission.  This supports a scenario where galaxies at higher
redshift have a lower neutral gas covering fraction, allowing both more
ionizing and \lya\ photons to escape directly to the observer, and that the observed increase
in the incidence at \lya\ emission at higher redshift may also trace
an increase in the ionizing escape fraction.  

\begin{figure*}[!t]
\epsscale{0.9}
\hspace{5mm}
\plotone{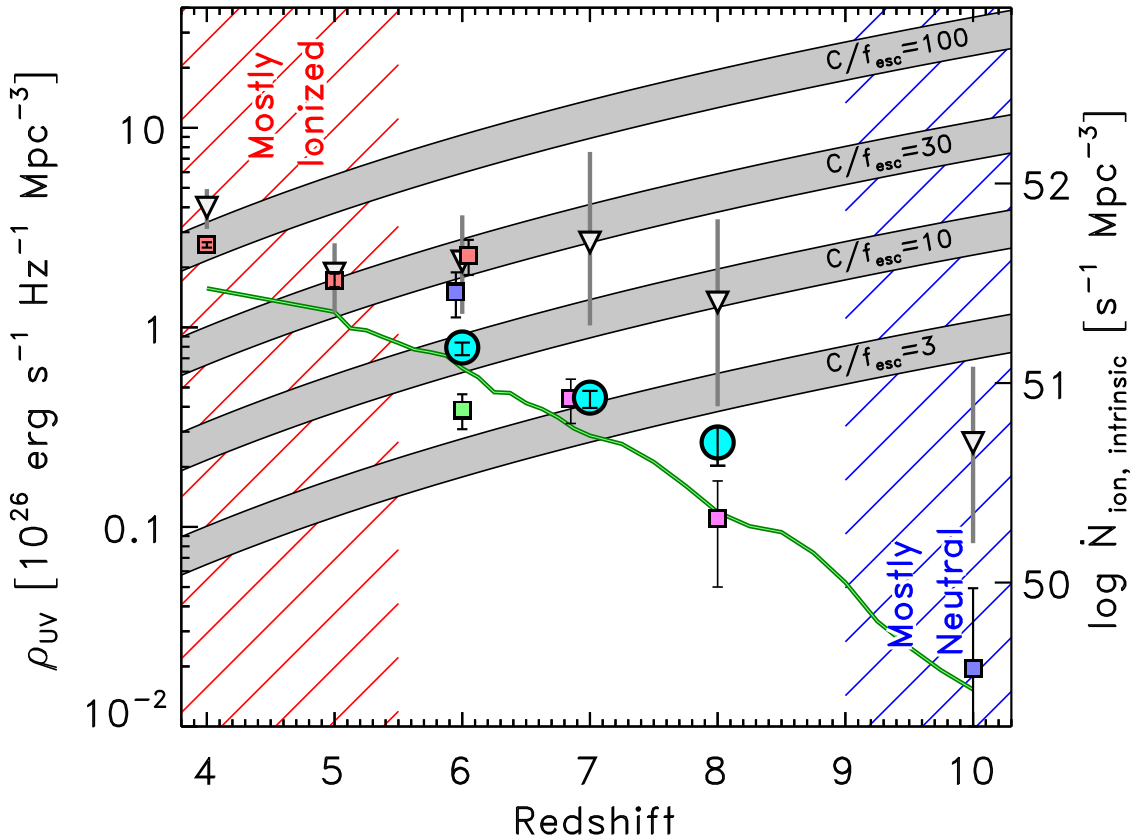}
\vspace{3mm}
\caption{The specific luminosity density ($\rho_{UV}$) versus
redshift, where our observed samples, corrected for incompleteness down to $M_{1500}$ $< -$18, are
plotted as the cyan circles.  Results from the literature are shown as
squares at $z =$ 4, 5 and 6 from the GOODS fields \citep[red;][]{giavalisco04}, $z =$ 6 from the
HUDF \citep[green][]{bunker04}, $z =$ 7 and 8 from the
HUDF \citep[purple;][]{finkelstein10a} and $z =$ 6 from GOODS+HUDF and
$z =$ 10 from the HUDF \citep[blue;][]{bouwens06,bouwens11c}.  All
squares have been adjusted to a limiting magnitude of $-$18 (with the
exception of the $z =$ 7 and 8 points, which were not corrected for incompleteness).  The inverted triangles denote the
integrated luminosity functions down to M$_{lim} = -$13 of \citet{bouwens07} at $z =$ 4, 5 and 6,
\citet{bouwens11} at $z =$ 7, \citet{bradley12} at $z =$ 8, and
\citet{bouwens11c} at $z =$ 10, where the gray error bar denotes the
uncertainty on these integrated values due to the uncertainty in the
Schechter function parameters (particularly the faint-end slope
$\alpha$ in the higher redshift bins).  The wide gray curves denote the
value of $\rho_{UV}$ needed to sustain a fully reionized IGM at a given
redshift, for a given ratio of the clumping factor C over the escape fraction of
ionizing photons f$_{esc}$ \citep{madau99}.  The width of these curves represent
changing stellar metallicities, from 0.02 $<$ Z/Z\sol $<$ 1.0.  The right-hand vertical axis shows
the corresponding intrinsic number density (i.e., prior to escape) of ionizing photons for a given
specific UV luminosity density, using the median of a range of ages
and metallicities, and assuming a constant star-formation history.
This axis can be multiplied by the reader's choice of f$_{esc}$.  The
green curve shows the predicted luminosity density for sources
brighter than M$_{1500} = -$18 from the hydrodynamic simulations of \citet{finlator11b}.}
\end{figure*} 
 
The reduced covering fraction may be due to
star-formation activity.  The specific star-formation rate
(sSFR; star-formation rate per unit mass) is remarkably constant from $z
\sim$ 4--7 \citep[e.g.,][]{stark09,gonzalez10}.  However, the typical
galaxy physical size at a given luminosity decreases towards higher redshift
\citep[e.g.,][]{ferguson04,oesch10b,ryan12}, thus the constant sSFR is
occurring over a smaller surface area, which can result in the
star-formation episode driving gas from the galaxies more easily \citep[e.g.,][]{kornei12}.  In
addition, the stellar masses of galaxies have been observed to
decrease with increasing redshift, reducing the gravitational
potential, also allowing more efficient
winds \citep[e.g.,][]{stark09,finkelstein10a,gonzalez10,labbe10b,mclure11},
plausibly decreasing the gas covering fraction, allowing a higher
escape fraction.  We note however that these measurements solely
constrain the stellar masses, not the total baryonic mass, which may
not evolve as strongly.
Finally, a higher escape fraction is also consistent with the observed
galaxy luminosity density and the
photoionization rate inferred from observations
of the \lya\ forest in high redshift quasars, which imply f$_{esc}$
$\sim$ 20--30\% \citep[see \S 4.5][]{bolton07}.

A number of theoretical studies have examined the escape of ionizing
photons, both as a function of redshift and halo mass.  For example, 
\citet{yajima11} investigated this quantity from $3 < z < 6$.  At $z
=$ 6, they found that the escape fraction was $>$ 10\% for all halos
more massive than 10$^{11}$ M\sol\ (which likely encompasses the
majority of the galaxies in our sample), with f$_{esc}$ rising towards
lower halo masses, for an overall average of 40\%.  \citet{razoumov10}
investigate this at higher redshifts, and found that typical escape
fractions in halos with masses from 10$^{10}$ -- 10$^{11}$ M\sol\ at
$z =$ 6--8 are $\sim$ 50--80\%.  The radiative transfer simulations of
\citet{haardt12} also find a relatively high escape fraction, and one
which rises with redshift.  They find that the luminosity weighted
escape fraction rises from 10 -- 30\% from $z =$ 6 $\rightarrow$ 8.  
Another recent study, \citet{conroy12} investigated the impact of runaway massive
stars on the overall escape fraction.  They found that by accounting for
the motions of these stars the escape fractions in lower mass
galaxies could rise, as the smaller sizes of lower mass galaxies would
allow massive stars to emerge from the gas disk prior to exploding.
They found that galaxies with halo masses less than 10$^{10}$ M\sol\
have escape fractions enhanced by 3--6$\times$ over models without
runaway stars.  Various other models
find that the escape fraction varies (either higher or lower) with
halo mass due to a variety of effects \citep[e.g.,][]{gnedin08, wise09}, including holes in the ISM from
supernovae \citep[e.g.,][]{yajima09}.  Thus, a relatively high escape fraction, and an escape
fraction that varies with halo mass and redshift appears justified, though it
remains to be verified observationally. 

In the following discussion, we will
use the value of C/f$_{esc} =$ 10 as our canonical model, which
represents a scenario of $C =$ 3 and f$_{esc} =$ 30\%, consistent with
the above results.  Should the reader prefer different values, we show
on Figure 3 critical curves for sustaining a fully reionized IGM for
four values of C/f$_{esc}$: 3, 10, 30 and 100.  In our discussion, we
will discuss the impact of a different escape fraction on our conclusions.

\subsection{Can the Observed Galaxies Reionize the Universe?}
In Figure 3, we show three previous results at $z =$ 6, all derived
solely using ACS optical imaging.  The first, from
\citet{giavalisco04} is based on a sample of 117 i$^{\prime}$-dropout
galaxies in the two GOODS fields.  Correcting their sample for
incompleteness down to M$_{UV} = -$19.3, they found $\rho_{UV} =$ 1.15 $\pm$
0.23 $\times$ 10$^{26}$ erg s$^{-1}$ Hz$^{-1}$ Mpc$^{-3}$.  To convert
this to our common magnitude limit of M$_{UV} = -$18, we use the
$z =$ 6 luminosity function of \citet{bouwens08}, finding a correction
factor of 2.0 (i.e., $\rho_{UV} [M < -18] =$ 2.3 $\pm$
0.46 $\times$ 10$^{26}$ erg s$^{-1}$ Hz$^{-1}$ Mpc$^{-3}$).  We note
that in their study, the existing depth in the z$^{\prime}$ data was
26.5; the data used in our present study is a full magnitude deeper.

The second observation, from \citet{bunker04} is based on a sample of 54 i$^{\prime}$-dropout
galaxies in the HUDF, finding $\rho_{UV} [M < -18.2] =$ 0.357 $\pm$
0.071 $\times$ 10$^{26}$ erg s$^{-1}$ Hz$^{-1}$ Mpc$^{-3}$; we apply a
factor of 1.08 to convert this to our common magnitude limit.
Finally, the third study, from \citet{bouwens06} uses
both of the aforementioned datasets, including also the HUDF parallel
fields, for a total sample of $\sim$500 galaxies, finding $\rho_{UV} [M < -17.5] =$ 1.77 $\pm$
0.45 $\times$ 10$^{26}$ erg s$^{-1}$ Hz$^{-1}$ Mpc$^{-3}$; we apply a
correction factor of 0.85 to these observations to convert to $\rho_{UV} [M < -18]$.

Interestingly, while the \citet{bunker04} survey goes deeper, down to M$_{UV}
\sim -$18.2, than the shallower GOODS data from \citet{giavalisco04},
one would have expected the former to yield a larger value of
$\rho_{UV}$ (prior to the correction), but the opposite is observed.
After we correct all three studies to a common magnitude limit, we
would expect them to be in agreement, which is not the case, as they
differ by more than a factor of five.
There are a few reasons why such a discrepancy is not surprising.  The
first is due to cosmic variance, as the \citet{bunker04} study only
probed the 11 arcmin$^2$ ACS data in the HUDF, while
\citet{giavalisco04} and \citet{bouwens06}
studies included the $\sim$ 300 arcmin$^2$ GOODS fields.  Secondly, as these
galaxies were selected only with ACS imaging in all three surveys, they were observed in only
a single band (z$^{\prime}$).  This can be
dangerous, as the likelihood of selecting a noise peak is greatly
enhanced when using a single band detection versus dual band (see
\citet{bouwens11c} and \citet{yan10} for a similar discrepancy in
single-band selected $z =$ 10 galaxy number counts; though the $z =$
10 disconnect is greatly enhanced due to the extreme faintness of such
sources).

The time is thus ripe to re-investigate the luminosity density at $z
=$ 6, using our up-to-date dataset.  Our current sample of $z =$ 6 galaxies is more robust, as our
selection over both the GOODS-S field and the HUDF09 fields covers both a large
volume, and a large dynamic range in object brightness.  Additionally,
the availability of the WFC3 data allows us to observe $z =$ 6
candidate galaxies in four bands, increasing the robustness of the
sample.  In particular, observations at four wavelengths throughout
the rest-frame UV yield much more robust estimates of M$_{UV}$ than
the single point previously available at $z =$ 6.  We find 302 $z =$ 6
galaxy candidates, or nearly double the size of
the \citet{giavalisco04} and \citet{bunker04}
samples combined (though less than \citet{bouwens06}; we expect this
to be rectified when we soon incorporate the incoming CANDELS data in
the GOODS-N field).  We measure $\rho_{UV} =$ 0.79 $\pm$ 0.06 $\times$10$^{26}$ erg s$^{-1}$
Hz$^{-1}$ Mpc$^{-3}$, which is in between the previously measured
values discussed above, with much smaller uncertainties.

As shown in Figure 3, the observable (i.e., $M_{1500} < -$18) $z =$ 6 galaxy population is
capable of sustaining reionization for our canonical model of a low
clumping factor and a moderate average escape fraction (C $=$ 3 and
f$_{esc} =$ 30\%).  Thus, in a change from a number of previous
studies \citep[e.g.,][]{yan04, bunker04}, it is not required
 to invoke galaxies fainter than our detection threshold to
complete reionization by $z =$ 6, due to both our more robust
measurement, as well as the updated assumptions about the clumping
factor.  Nor do we need to invoke an escape fraction of unity, as
f$_{esc} \sim$ 30\% on average is sufficient (if C $\approx$ 3).  Finally, we reiterate this is also
independent of assumptions about the form and faint-end cutoff of the
luminosity function.  We note that our inclusion of deeper
$z^{\prime}$ imaging, multiple fields, as well as near-infrared
imaging yields a more robust estimate of the luminosity density, in
contrast to the previous observations, which differed by a factor of
$>$ 5$\times$.

Different conclusions are reached at $z =$ 7--8.  As shown in Figure 3, the specific UV luminosity
density experiences a steady decline with redshift from $z =$ 4 -- 8.
Combined with the denser IGM at higher redshift, necessitating a
higher $\rho_{UV}$ to compensate for the shorter recombination time,
the observable galaxy population at $z =$ 7--8 falls short of what is
needed to sustain a fully reionized IGM, unless the escape fraction is unity
(which is unlikely to be the case for all galaxies).  A similar result is seen in the
simulations of \citet{salvaterra11}, as shown in their Figure 8.  They
find that galaxies observable with {\it HST} can sustain a reionized
IGM if f$_{esc} =$ 20\% and C $=$ 5 at $z =$ 6, with C $<$ 2 (or a
higher escape fraction) necessary at $z \geq$ 7 to maintain
reionization with the observable galaxies.  However, their simulation
also probes down to M$_{UV} = -$15, and at that level they find that
the observable galaxies with {\it HST} only account for $\sim$20\% of
the total number of escaping ionizing photons.  

At $z =$ 7, we find that the observed galaxies only account for
$\sim$50\% of the required UV luminosity density inferred to sustain
reionization.  As we discuss later in \S 4.7, observations reaching
$\sim$ 1 magnitude fainter may yield enough galaxies (assuming
the published luminosity functions) to account for a fully reionized
IGM {\it if} C/f$_{esc} =$ 10, similar to conclusions at this redshift
from \citet{robertson10}.  We explore the possible contribution from
galaxies below our detection threshold in \S 4.4, and examine possible
changing escape fractions in \S 4.5.

Our observations from only the observed galaxy population are consistent with independent evidence that
the universe was not completely reionized at $z \geq$ 7.  Studies of both \lya\ emission
\citep[e.g.,][]{stark10,stark11} and quasar near-zones \citep[e.g.,][]{fan06} at $z \sim$ 6 
yield results consistent with a negligible neutral fraction, implying
that reionization was complete by $z =$ 6.  However, recent
spectroscopic followup of $z \sim$ 7 galaxies has yielded a much lower
success rate of \lya\ emission detection than there should have been
given the \lya\ EW distributions at $z =$ 4 -- 6
\citep{stark10,fontana10,pentericci11,ono12,schenker12,treu12}, implying
neutral fractions as high as $\sim$50\%.  A similar decrease in the
\lya\ luminosity function has been measured from $z =$ 5.7 to $z =$
6.6 \citep[e.g.,][]{ouchi10,hu10}, though see also \citet{malhotra04}.  Followup high-resolution
spectroscopic study of the near-zone around the recently discovered $z
=$ 7 quasar ULAS J1120+0641 \citep{mortlock11} yields evidence for a
neutral fraction $>$ 10\%.  We note that the \lya\ results are also
dependent on the neutral gas distribution within the galaxy
\citep[e.g., see \S 6.3 of][]{finkelstein11c}, and the quasar results
are based on a single sightline, which could be biased by a nearby
cloud of neutral gas \citep[e.g.,][]{bolton11}, thus more data are
needed to conclusively show that the neutral fraction is indeed rising.
In any case, if our assumptions of a low clumping factor and a moderate escape
fraction are reasonably correct, than the combination of galaxy,
quasar and spectroscopic observations point to an end to
reionization by $z =$ 6, but not earlier than $z \sim$ 6.5.  This then
implies that through some combination of limiting magnitude and
luminosity dependent escape fractions, the observed galaxy populations are an excellent tracer
of the ionizing photon background.  

In order to place these results in context, we examine the recent
study of \citet{kuhlen12}, who combined the {\it WMAP} and SPT results with the observed
luminosity function evolution (extrapolated to higher redshift), and
$z <$ 6 \lya\ forest measurements to investigate a concordance model
of reionization.  They find that the available data
predict that the universe was $\sim$50\% ionized at $z \sim$ 10.  They
note that, given the (less conservative) $\Delta z <$ 4.4 constraint from the SPT
measurements \citep{zahn11}, scenarios invoking a very high
contribution from very faint galaxies will result in a reionization
history which is too extended.  However, recent work by
\citep{mesinger12} shows that there likely do
exist correlations between the thermal SZ power and the cosmic
infrared background; in the presence of these correlations, the
constraint on the duration of reionization degrades to $\Delta z <$
7.9 \citep{zahn11,mesinger12}.  As shown in Figure 3, if the IGM is
truly $\sim$ 50\ ionized at $z \sim$ 10, then there will need to be a
significant contribution from faint galaxies, necessitating a longer
duration of reionization.  These conclusions thus stand to be revisited when a better
understanding of this effect and its possible
dependance on correlations are obtained.

We note that our quoted rest-UV luminosities are not corrected for dust attenuation.  In \citet{finkelstein11c}, we
examined the average dust attenuation as a function of redshift,
finding that $z =$ 6 galaxies appeared to contain little dust, and $z
=$ 7 galaxies appeared to be dust free (due to only two detected
fluxes, robust estimates of the extinction at $z =$ 8 were difficult, though
they were consistent with little-to-no dust).  Using the values
of the UV spectral slope $\beta$ from that work, and the conversion
from $\beta$ to the attenuation at 1600 \AA\ from \citet[][see
discussion in Finkelstein et al.\ 2011 for the applicability of this
relation at high redshift]{meurer99}, we find that A$_{1600} =$ 0.31 mag
at $z =$ 6, and A$_{1600} =$ 0 mag at $z =$ 7 and 8.  Thus, accounting
for dust does not change the above conclusions at $z =$ 7 and 8, while
the observed specific luminosity density may in fact be up to 30\% higher
at $z =$ 6.  However, any dust present will likely also attenuate the
ionizing photons, though the dust attenuation curves are not well
sampled at such blue wavelengths \citet[e.g.,][]{calzetti94}.  In light of
this, we conclude that using the observed luminosity density is
appropriate, assuming that the ionizing radiation is not
significantly more (or less) attenuated than the rest-frame 1500 \AA\ light.

Finally, in Figure 3 we show in green the predicted luminosity density for
galaxies at M$_{1500} < -$18, derived from a radiation hydrodynamic
simulation that incorporates both galactic outflows and a self-consistently 
grown ionizing background as in \citet{finlator11b}, while subtending a 12 h$^{-1}$ Mpc 
volume at the same resolution.  At all redshifts, these predicted values fall
short of our observed luminosity densities.  Some of this discrepancy
may be due to volume limitations, but as we show below, the
sub-$L^{\ast}$  galaxies are the dominant contributors, hence they
should be reasonably well-represented.  This discrepancy could indicate that 
a higher star-formation efficiency in the models is warranted, perhaps via 
weaker outflows.  It also relaxes the requirements for the ionizing escape fraction.  In particular, the 
\citet{finlator11b} models required an escape fraction of 50\% in order to complete
reionization by $z =$ 6; weakening feedback by a factor of 2--3 would suppress 
the required escape fraction by the same factor, bringing it closer to the
observed values at lower redshift \citep{shapley06, siana10,
  nestor11}, as well as those we find below in \S4.5.

\subsection{Contribution of Different Luminosity Ranges}

We now investigate how the different luminosity ranges of our
observed galaxy sample contribute to the total specific UV luminosity
density. We reiterate that this is independent of any luminosity
function parameterization (with the exception of the definition of the
characteristic luminosity $L^{\ast}$).  In Figure 4, we show the fraction of the critical value of
$\rho_{UV}$ necessary for sustaining a fully reionized IGM reached by our observed population at each redshift, for
all galaxies (large circles), and for galaxies with luminosities
greater than 0.2, 0.5 and 1 $\times$ $L^{\ast}$  (small colored
symbols), where the appropriate value of $L^{\ast}$  for each redshift
is used.

\begin{figure}[!t]
\epsscale{1.2}
\hspace{-10mm}
\plotone{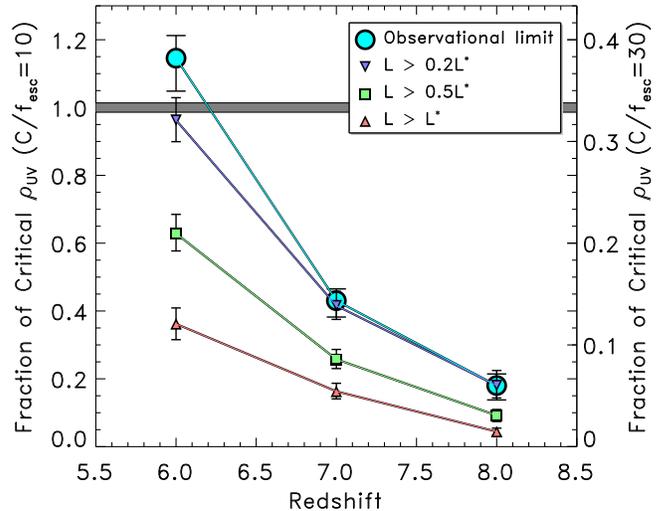}
\vspace{-2mm}
\caption{The fraction of the critical specific UV luminosity density
  necessary for sustaining reionization (denoted by the gray bar for
  C/f$_{esc} =$ 10) obtained from the
  (incompleteness-corrected) observed
  galaxy population (large cyan points) as a function of redshift,
  assuming C/f$_{esc} =$ 10 (left vertical axis) and 30 (right
  vertical axis).  The blue  inverted triangles, green squares, and red triangles show the
  critical fraction for galaxies with luminosities greater than 0.2,
  0.5 and 1 $\times$ the characteristic luminosity $L^{\ast}$ ,
  respectively.  At $z =$ 6, where the observed population can reionize
the universe, this figure highlights that while galaxies below our
detection limit are not necessary, it is the fainter portion of the
observed population which is dominating $\rho_{UV}$, as galaxies with
luminosities greater than 0.2$L^{\ast}$  contribute $\sim$2.6$\times$
more than L $>$ $L^{\ast}$  galaxies.  This is even more so the case at
$z =$ 7 and 8, though at these redshifts the entire observed population is short by a
factor of $\sim$2.5--3 of the ionizing photons necessary to sustain
reionization for C/f$_{esc} =$ 10.}
\end{figure} 

At $z =$ 6, this figure reinforces our observation that the observable
population at $z =$ 6 appears sufficient to maintain an ionized IGM,
while those at $z =$ 7 and 8 are not.  At $z =$ 6, the ratio of the specific UV
luminosity density from galaxies with L $>$ 0.2$L^{\ast}$  to that of
galaxies with L $>$ $L^{\ast}$  is a factor of $\sim$2.6.  This
highlights that it is very much the sub-$L^{\ast}$  galaxies which are
dominating the reionization photon budget.  Additionally, galaxies with 0.2$L^{\ast}$ 
$<$ L $<$ 0.5$L^{\ast}$  contribute a roughly equal amount to
reionization as those with L $>$ $L^{\ast}$ .  These ratios are similar
at $z =$ 7.  

At $z =$ 8, the results are even more tilted in favor of the fainter
galaxies, with the ratio of $\rho_{UV}$ from galaxies with L $>$ 0.2$L^{\ast}$  to that of
galaxies with L $>$ $L^{\ast}$  at a factor of $\sim$4.8, and
galaxies at 0.2L$^{\ast} <$ L $<$ 0.5$L^{\ast}$  contributing double
the number of UV (and plausibly ionizing) photons as the L $>$
$L^{\ast}$  galaxies (though both woefully short of the critical value
of $\rho_{UV}$).  Our observations thus imply that the faint-end
slope of the luminosity function at $z \geq$ 6 is steep, and that the
slope is becoming even steeper as $z \rightarrow$ 8 \citep[in
agreement with previous studies; e.g.,][]{bouwens11d,oesch12,bradley12}.

\subsection{Galaxies Below our Detection Threshold}

Although our observed population appears consistent with multiple
lines of evidence pointing to the completion of reionization by, but
not too long before, $z =$ 6, our conclusions are dominated by
assumptions about the escape fraction of ionizing photons.  Thus, it
may be that fainter galaxies do play a large role.  Although we will
not observe them until the launch of the {\it James Webb Space
  Telescope} ({\it JWST}), it is highly likely that there
do exist galaxies below our observational limit.  However, this
introduces another uncertainty, in that we do not know how far down
the luminosity function galaxies are able to form.

At some limiting dark matter halo mass, the
gas within a halo will no longer be able to self-shield against
an ionizing background from the sources which began reionization.  In
such halos star formation will be suppressed, effectively cutting off
the UV luminosity function at a limiting magnitude M$_{lim}$
\citep[e.g.,][]{shapiro94,thoul96,gnedin00,dijkstra04,okamoto08,finlator11b},
though in the case of inhomogeneous reionization, very small halos may
still contribute a significant number of ionizing photons before they
are quenched \citep{finlator11b}.
This has not yet been seen observationally,
as there is no evidence for a turnover in the luminosity
function \citep[e.g.,][]{bouwens07, bouwens11},
which implies that such a cutoff occurs at M$_{lim} > -$16.5 (the
limiting absolute magnitude for $z =$ 4 galaxies in the HUDF).
Such a turnover has also not been seen in large cosmological
simulations.  For example, the models of \citet{finlator11b} find
an increasing luminosity function down to their resolution limit, which
corresponds to M$_{lim} = -$13.  Interestingly, a recent study of high redshift gamma
ray burst (GRB) host galaxies by \citet{trenti12} finds that the
non-detection of five $z >$ 5 GRB hosts implies that galaxies fainter
than $M_{UV} = -$15 were likely present at $z >$ 5 \citep[see also][]{tanvir12}.

The inverted triangles in Figure 3 show the specific UV luminosity
density if we take the luminosity functions from the literature, and
integrate them down to M$_{lim} = -$13
\citep{bouwens07,bouwens11c,bouwens11,bradley12}.  The uncertainties
on these values were derived via 10$^{3}$ Monte Carlo simulations,
where in each simulation each Schechter function parameter was
modified by a Gaussian random number multiplied by the associated
parameter uncertainty. 

\begin{figure*}[!th]
\epsscale{1.2}
\hspace{-7mm}
\plottwo{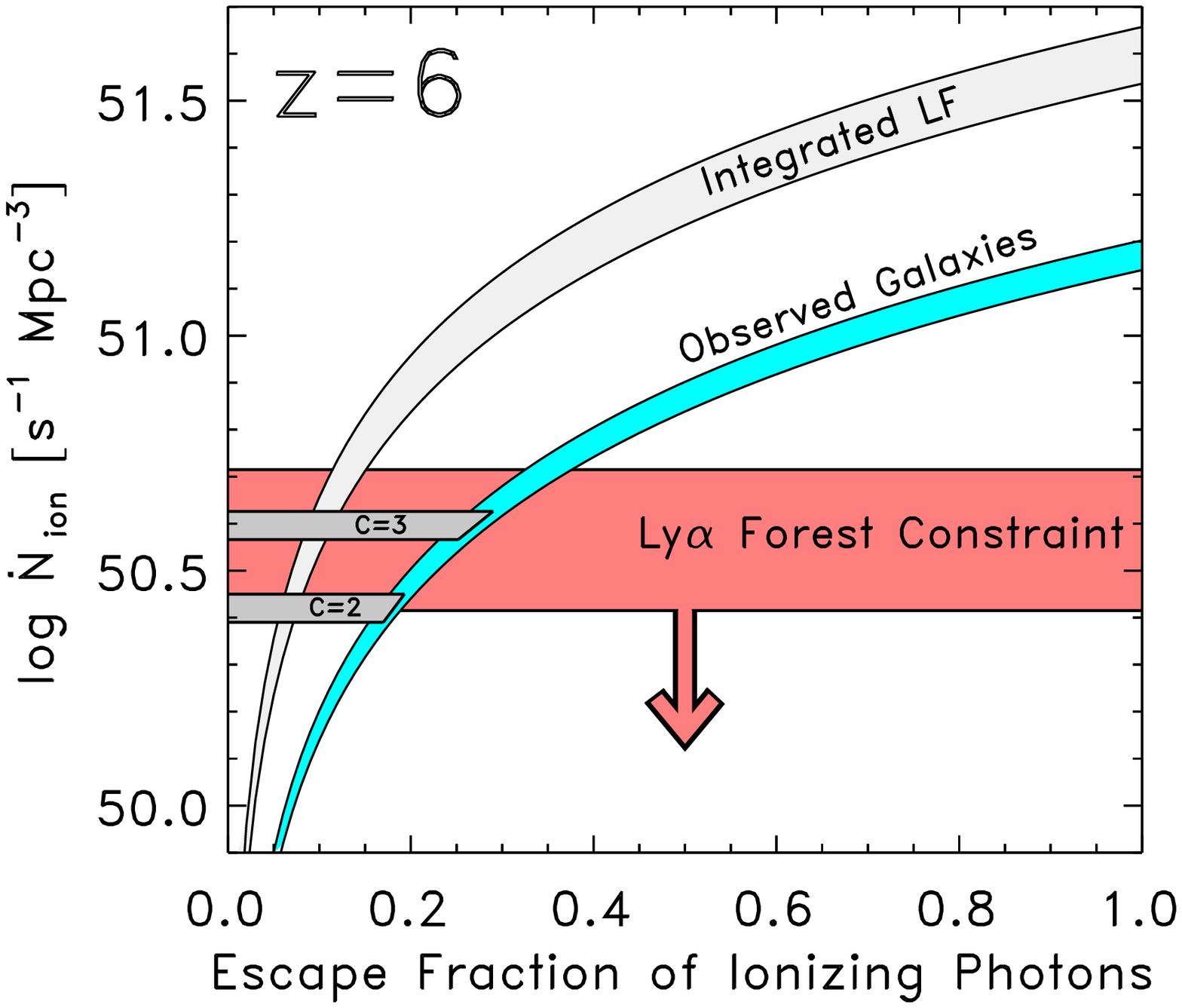}{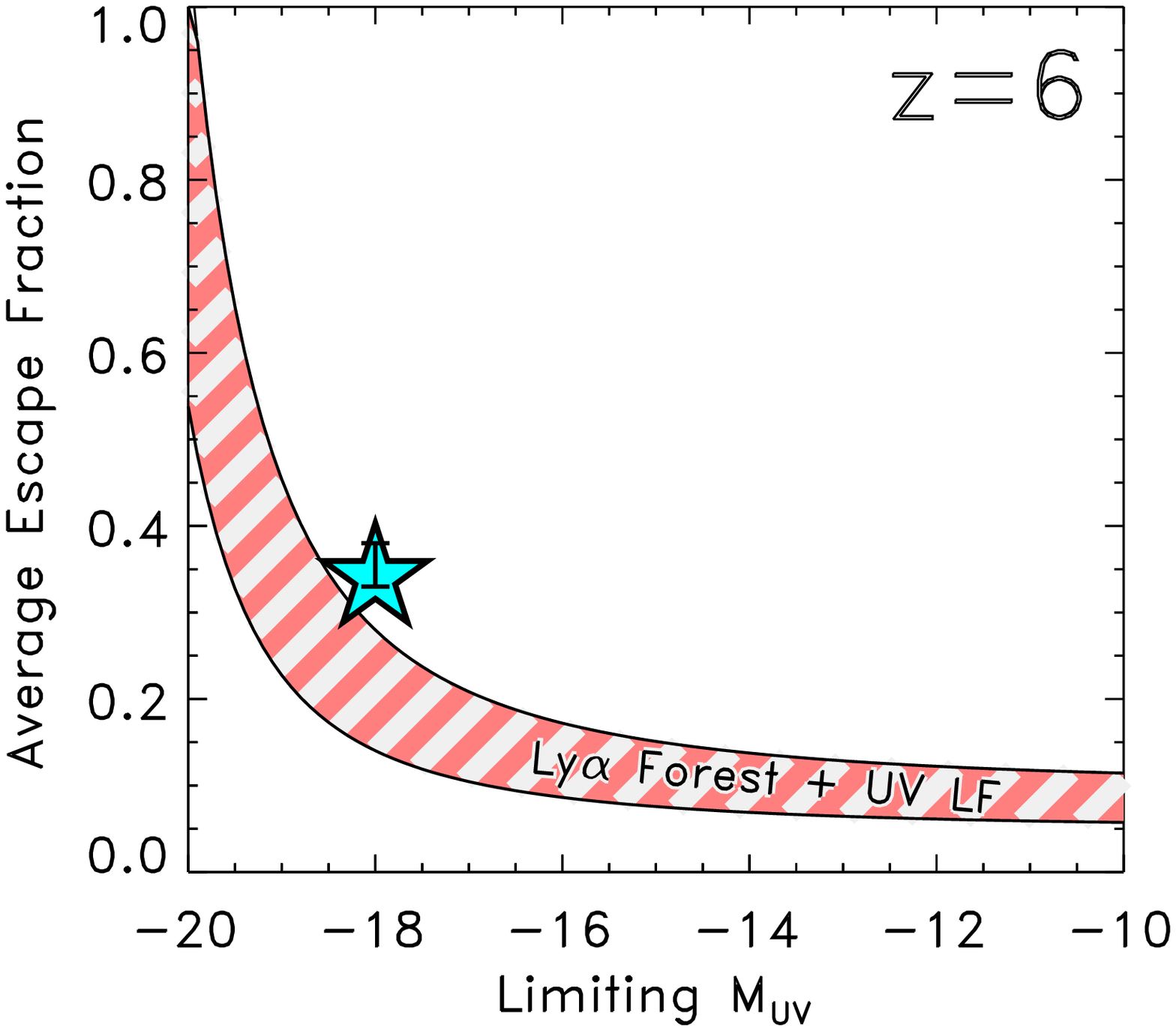}
\vspace{-2mm}
\caption{Left) The inferred emission rate of ionizing photons from our
  (incompleteness-corrected) observed sample of $z =$ 6 galaxies (cyan; the width of curve denotes the 68\%
  confidence range), as well as from the $z =$ 6 luminosity
  function of \citet{bouwens08} integrated down to M$_{UV} = -$10
  (left side of light-gray curve) and  $-$15 (right side).
  Both quantities are plotted as a function of the escape fraction of ionizing photons.  The red
  region denotes the constraints on this quantity at $z =$ 6 from the Ly$\alpha$
  forest, with the lower and upper edges representing the 1 and
  2$\sigma$ constraints from \citet{bolton07}, respectively.  The dark gray bands denote the emission rate of
  ionizing photons necessary to sustain reionization for a given value
  of the clumping factor.  
Right) The striped curve shows the allowed average escape fraction of ionizing photons from
galaxies as a function of the limiting UV absolute magnitude, assuming
the $z =$ 6 luminosity function of \citet{bouwens07}, based on
the 2$\sigma$ constraints from the \lya\ forest.  The cyan star shows the
allowed escape fraction from our observations with its associated uncertainty.  If the galaxy
  luminosity function turns over shortly faintward of our
  observations, then the average escape fraction of ionizing photons is
  constrained to be f$_{esc} <$ 34\% (17\%) at 2$\sigma$ (1$\sigma$).  If the
  luminosity function extends down to  M$_{UV} = -$13, then this
  constraint tightens to f$_{esc} <$ 13\% (6\%) at 2$\sigma$
  (1$\sigma$).  These escape fractions can still reionize the IGM if
  the clumping factor is $\sim$ 2--3.}
\vspace{5mm}
\end{figure*} 

We note that in these simulations, we have used the published
uncertainties on each Schechter function parameter; thus, they were
free to vary independent of each other.  However, these parameters, 
particularly the characteristic magnitude
$M^{\ast}$, and the faint-end slope $\alpha$, are 
correlated, thus the uncertainty on the integrated luminosity
functions may be smaller than those derived here.  
If the published luminosity functions accurately represent the true
galaxy population, and the population does indeed extend this faint, than the
existing galaxy population at $z =$ 7 and $z =$ 8 can sustain
reionization for C/f$_{esc} =$ 10.  However, both luminosity functions are highly uncertain
at this point, especially at $z =$ 8, where the error bar stretches
well below the value necessary to sustain reionization.  These
luminosity function uncertainties, combined with the uncertainty on
the value of M$_{lim}$, render the contribution of the unseen faint
galaxies to reionization difficult to assess.

\subsection{Constraints on the Escape Fraction}

As discussed above, if the escape fraction is modestly high, at
$\sim$30\%, then the observable galaxy population at $z =$ 6 can
sustain a fully reionized IGM.  However, it is highly probable that
galaxies significantly fainter than this limit exist, given results from a wide variety of cosmological simulations
\citep[e.g.,][]{trenti10,salvaterra11,finlator11b,jaacks12}, and plausible,
though scant, observational evidence \citep[e.g.,][]{trenti12}.

In light of the likelihood that the galaxy luminosity function extends
fainter than we can observe, it is instructive to examine the effect
of our choice of 30\% for the escape fraction in the above discussion.  As shown in Figure 3,
our observed galaxy population will yield just enough photons to
complete reionization at $z =$ 6 if the escape fraction is 30\%.  If
the luminosity function extends significantly fainter than M$_{UV} =
-$18, and the escape fraction is on average 30\% for all galaxies,
then more ionizing photons will be produced than are necessary to
complete reionization.  In principle, this is not a problem, as once
the IGM is reionized, it will stay ionized as long as the critical
number of ionizing photons are being produced.

However, there is one observational constraint we have not yet strongly
considered, which is the measurement of the ionizing emissivity from
observations of the Ly$\alpha$ forest opacity.  \citet{bolton07}
studied this at $z =$ 6 combining simulations with high-resolution spectroscopic
observations of $z >$ 6 quasars, and found that only 1.5 -- 3 ionizing
photons per hydrogen atom were created, leading them to believe that
reionization was ``photon-starved.''  At $z =$ 6, they calculated a
1$\sigma$ upper limit on the emission rate of ionizing photons per
comoving Mpc of $<$ 2.6 $\times$ 10$^{50}$ s$^{-1}$ Mpc$^{-3}$.  This
result, though based on a relatively small number of sight-lines, implies an upper limit to the number
of ionizing photons created at $z =$ 6.

In Figure 5, we compare this constraint from the Ly$\alpha$ forest to
our observations, for a variety of escape fractions.  We convert our
observed UV luminosity density into a comoving rate of ionizing
photons using the same set of stellar population model assumptions
used to derive $\epsilon_{53}$ in \S 3.4.  Assuming $Z =0.2Z$\sol, we
find \.{N}$_{ion}$ = 1.9 $\times$ 10$^{25} \times \rho_{UV} \times$ f$_{esc}$.  This assumes
a star-formation history that is in equilibrium (i.e., massive O stars
die as fast as they are formed).  We use this to show in Figure 5 the
inferred emission rate of ionizing photons from both our observed
galaxies (in cyan), and the $z =$ 6 luminosity function of
\citet{bouwens08} integrated down to M$_{UV} = -$10 and $-$15 (light
gray band, where the left and right edges of the band denote M$_{lim}$
= $-$10 and $-$15, respectively).

If M$_{lim} \sim -$18, an escape fraction of 30\% is allowed within the
2$\sigma$ upper limit on the Ly$\alpha$ forest constraint.  With this
constraint, the 2$\sigma$ (1$\sigma$) upper limit on the escape fraction is $<$
34\% (17\%).  As shown by the light gray curve, if the luminosity function extends fainter
than our observations, then the escape fraction \emph{must} decrease,
otherwise the Ly$\alpha$ forest constraints will be violated.  Using
the limit of M$_{UV} = -$13 used above in \S 4.4, we find that
accounting for galaxies fainter than our detection limit the average
escape fraction must be $<$ 13\% (6\%) at 2$\sigma$ (1$\sigma$).  We
note that the 2$\sigma$ limit on f$_{esc}$ in both cases (observed and integrated luminosity function) are
consistent with sustaining reionization if the clumping factor C $=$ 3.  If future
observations constrain the ionizing photon production rate to be
closer to the 1$\sigma$ limit, then the galaxy population can only
sustain reionization if C $\leq$ 2. 

Given the current observations, we can thus say with 95\% confidence
that the average escape fraction is $<$ 34\%.  Future observations
with {\it JWST} will push the luminosity functions several magnitudes
lower, and thus put more robust constraints on this quantity.  We
however note that current observations constraining the ionizing background at
$z =$ 6 are due to a relative low number of individual sight-lines to
distant quasars.  Additionally, these quasars reside in biased
regions, thus their spectra may not represent a true probe of the mean
IGM neutral fraction.  Ongoing and future studies such as UKIDDS and VIKING will yield
much greater probes of the ionizing background at $z =$ 6--7, which can
then be turned into stronger constraints on the escape fraction of
ionizing photons from the $z =$ 6 galaxy population.

Finally, while the above discussion assumes the escape fraction
represents an average of all galaxies, this does not have to be the
case.  It is more accurately a luminosity-weighted escape fraction; essentially, this
quantity tells us what fraction of the total number of ionizing
photons produced in all galaxies escape into the IGM.  It may also be the case that
the escape fraction is time dependent; ionizing photons may only
escape from galaxies due to specific temporal events, such as
interactions or starbursts.  Our constraints limit the total number of ionizing photons which escape
from all galaxies, which can be consistent with a time,
luminosity or mass dependance, or no dependance at all.

\subsection{Constraints on the Reionization History}
As we move into the reionization epoch, and the ionized fraction in the universe becomes 
less than unity, one can use Equation 2 to calculate the luminosity density
necessary to maintain a given ionized fraction by varying 
the volume ionized fraction $x_{HII}$.  We can thus
indirectly estimate $x_{HII}$ as a function of redshift by dividing
our observations by the critical value of $\rho_{UV}$ necessary to
sustain a fully ionized IGM as a function of
redshift.  This approach is a valid approximation to semi-analytic modeling
\citep{madau99} because the IGM recombination time remains less than
the Hubble time for $C =$ 3 at $z >$ 6.
This is subject to the same assumptions as the above analysis, including:
\begin{itemize}
\item Galaxies are the only source of ionizing photons in the $z \geq$
  6 universe.  While other sources (i.e., the rare quasar) may
  exist, their overall contribution is likely negligible.
\item The faint-end cutoff of the luminosity function.  As above, we
  will consider two scenarios: 1) Only galaxies above the
  observational limit of M$_{UV} = -$18, and 2) The published
  luminosity functions integrated down to $M_{UV} = -$13.
\item The clumping factor of ionized hydrogen, which we take to be $C$
  $=$ 3.
\item The escape fraction of ionizing photons; per \S 4.5, we assume here f$_{esc} =$ 30\% when
  considering only the observable galaxies, and f$_{esc} =$ 10\% when
considering the integrated luminosity functions.  Both scenarios are
consistent with the Ly$\alpha$ forest constraints, and both can sustain
a fully reionized IGM at $z =$ 6 if C $\lesssim$ 3.  
\item We assume that the ratio C/f$_{esc}$ does not evolve over the redshifts we study.
\end{itemize}

We plot the results of this analysis in Figure 6.  Our
incompleteness-corrected measurement from the observed galaxies
(assuming f$_{esc} =$ 30\%) is
consistent with a volume ionized fraction of 1.0 at $z =$ 6, $\sim$0.4
at $z =$ 7, and $\sim$0.2 at $z =$ 8.  When we consider the integrated
luminosity function, with f$_{esc} =$ 10\%, we find a much wider range
of possible volume ionized fractions.  We note that while the
UV luminosity density from the integrated luminosity function will always
be larger than that for galaxies above our observational limit, here
we assume two different escape fractions for the two magnitude limits,
thus their ionizing luminosity densities are comparable.  At $z =$ 6, the integrated
luminosity functions are consistent with ionized fractions of $\sim$
0.7 -- 1.0.  At $z =$ 7 and 8, this is reduced to 0.3 -- 1.0, and 0.1
-- 0.95, respectively.  The very wide range of possible ionized
fractions from the integrated luminosity functions are due to the
increasingly larger uncertainties in the Schechter function parameters
as one pushes to higher redshift.  

Also on Figure 6, we plot a number of results from the literature,
including constraints on the volume ionized fraction from quasars at
$z \leq$ 6 \citep{fan06} and $z =$ 7 \citep{bolton11}, {\it WMAP}
\citep{komatsu11}, \lya\ emitter luminosity functions \citep{ouchi10},
and \lya\ spectroscopic studies \citep{pentericci11, schenker12,
  ono12}.  Both our observed galaxy population, and the integrated
luminosity functions, are consistent with this wide and complementary
range of reionization probes.  The only exception is the single known
quasar at $z =$ 7, which has a near-zone size consistent with an
ionized fraction of $\lesssim$ 90\%, and appears slightly inconsistent with our
observed galaxy population (though consistent with the integrated
luminosity functions).  However, this is but a single sightline
to the only currently known $z \geq$ 7 quasar, thus many more
sight-lines are needed to truly constrain the IGM from this type of
probe.

\begin{figure*}[!t]
\epsscale{0.8}
\hspace{-5mm}
\plotone{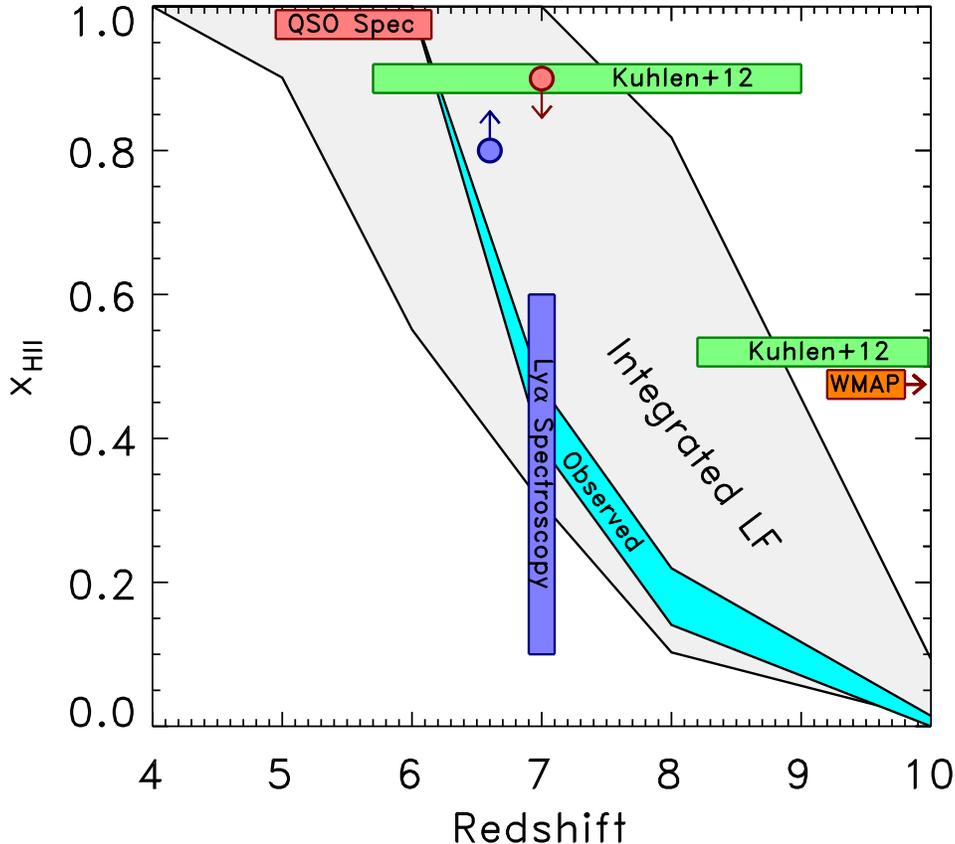}
\caption{The volume ionized fraction, x$_{HII}$, of the IGM which can
  be sustained given the observed UV luminosity densities (and our
  assumptions on C and f$_{esc}$) as a function of
  redshift.  The cyan and gray curves denote the value of x$_{HII}$
  inferred from comparing our observed galaxies, and the integrated
  luminosity functions, respectively, to the critical luminosity
  density from Equation 2.  We have assumed $C/f_{esc} =$ 10 when only considering the
observable galaxies, and $C/f_{esc} =$ 30 when considering
the integrated luminosity functions (both of which are
consistent with limits from the $z =$ 6 \lya\ forest presented in \S
4.5, for $C =$ 3).  We note that while the luminosity density from the integrated luminosity functions will always
be larger than that for galaxies above our observational limit, here
the two different escape fractions for the two magnitude limits result
in their ionizing luminosity densities being comparable.  
We assume that galaxies are the only significant sources
of ionizing photons in this epoch.  We plot constraints on x$_{HII}$
from spectroscopy of quasars at $z <$ 6 from \citet{fan06} and at $z
=$ 7 from \citet{bolton11} in red.  The blue circle denotes constraints on
x$_{HII}$ from the evolution in the \lya\ luminosity function from $z
=$ 5.7 to 6.6 from \citet{ouchi10}, while the blue bar denotes the
range of x$_{HII}$ values inferred from the $z =$ 7 follow-up \lya\
spectroscopic studies of \citet{pentericci11, schenker12, ono12}.  The
instantaneous redshift for reionization from {\it WMAP} (10.6 $\pm$ 1.2) is
indicated by the orange rectangle.  The derived 50\% and 90\%
x$_{HII}$ redshifts from the study of \citet{kuhlen12} are shown in
green, where the width of the green bars denote the 1$\sigma$
uncertainty (from Figure 9 of Kuhlen et al.).  Our observations, which
are consistent with a wide range of other, complementary, approches,
find a picture where the universe is fully ionized by $z =$ 6, with
the neutral fraction becoming non-negligible at $z \gtrsim$ 7.  This
is independent of whether we consider only the observable galaxies, or
the integrated luminosity functions, each with their own permitted
escape fraction.  These observations are in moderate tension with the
concordance models of Kuhlen et al.\ at $z >$ 9, as these models
prefer an earlier redshift for x$_{HII} =$ 50\%.  This may be resolved
with future deep observations, which will provide much better
constraints on the $z \geq$ 8 luminosity functions, and show if the
strong drop-off in luminosity density at $z >$ 8, currently inferred
from a single source, is accurate.
}
\vspace{5mm}
\end{figure*} 

Our new observations combined with the previous, complementary, probes
of the ionization state of the IGM show a consistent reionization
picture, where the ionized fraction in
the IGM is essentially unity at $z \leq$ 6.  At $z \sim$ 7, many lines of
evidence indicate a non-negligible neutral fraction, including our
observed galaxy population.  The integrated luminosity function
prefers an ionized fraction $<$ unity, but it is consistent with a
fully ionized IGM at 1$\sigma$.  This is not the case at $z =$ 8,
where galaxies are the only probe we currently have, and they seem to
indicate an ionized fraction of $<$ 80--90\% (unless the escape fraction
in all galaxies is unity, which would indicate extreme evolution from
$z =$ 6, and is unlikely; see \S 4.5).  Unfortunately, more robust constraints on
the evolution of x$_{HII}$ from the integrated luminosity functions
are difficult due to the uncertainty in the Schechter function parameters.
We note that we have used the uncertainties on each Schechter function parameter to compute the
uncertainty in the integrated luminosity function (see \S 4.4 for more
details).  These parameters, particularly the characteristic magnitude
$M^{\ast}$, and the faint-end slope $\alpha$, are known to be
correlated, thus the uncertainty on the integrated luminosity
functions may be smaller than those derived here.  In any case, at $z
\geq$ 7, the faint-end slope is not well constrained, which is
responsible for the bulk of the uncertainty on x$_{HII}$.  More robust
luminosity functions in this epoch will help decrease the uncertainty
on x$_{HII}$ from galaxy measurements.

Finally on Figure 6, we also show the concordance models of
\citet{kuhlen12}, which present a picture of the reionization history
of the universe consistent with the \lya\ forest, {\it WMAP} optical depth,
and the kSZ effect.  They also fold in the observed galaxy luminosity
functions, though in this case they assume that the luminosity
function parameters evolve smoothly with redshift; we do not know if
this is truly the case, particularly at $z >$ 8, beyond which the strong
drop-off in the observed luminosity density is intriguing.  We plot
their results for the redshift of a 50\% and 90\% ionized fraction.
Comparing our inferred values of x$_{HII}$ from galaxies
with their conclusions, we find general agreement at $z <$ 8, in that
they find that reionization completes by $z =$ 6, and that it is 90\%
complete at $z =$ 7.5 $\pm$ 1.5.  

\begin{figure*}[!t]
\epsscale{0.9}
\hspace{-10mm}
\plotone{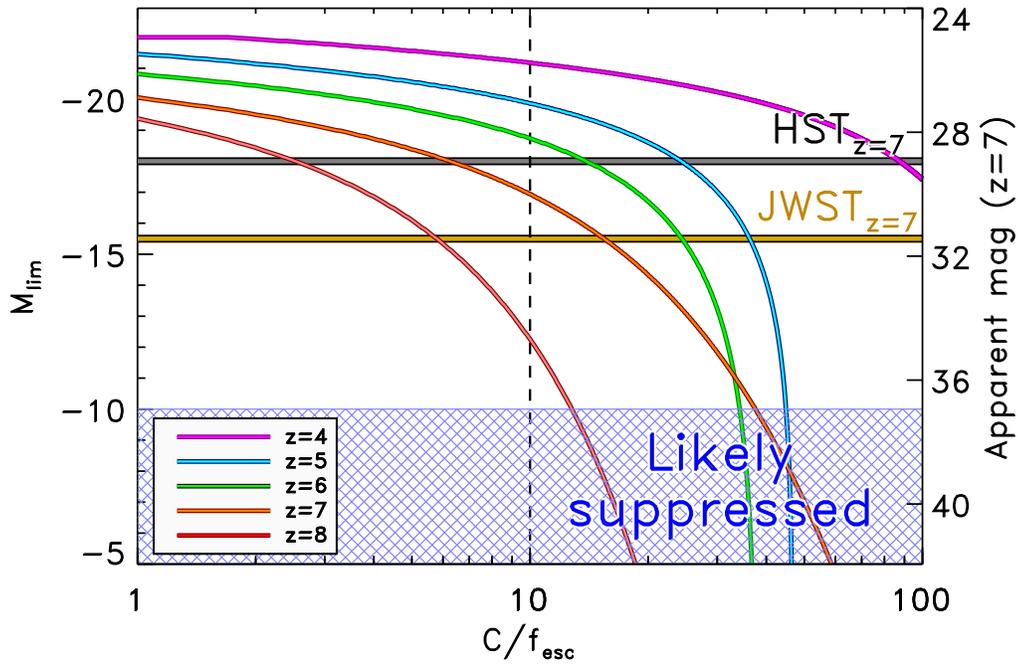}
\vspace{-2mm}
\caption{The colored curves show the limiting UV
  absolute magnitude for different redshifts necessary to sustain
  reionization for a given ratio between the IGM clumping factor and
the ionizing escape fraction.  The hatched region denotes the
magnitude range where star formation is likely to be suppressed via
photo-heating from the sources which began the process of reionization
at higher redshift.  The dashed line shows our canonical
model of a clumping factor of 3 and f$_{esc} =$ 30\%, or C/f$_{esc} =$
10, which is a reasonable assumption for the conditions in the
universe at $z \geq$ 6.  The horizontal gray bar denotes the limiting
UV magnitude of {\it HST} observations of the HUDF at $z =$ 7 (m$_{AB}
=$ 29).  A hypothetical JWST deep
field will reach 2.5 mag deeper, to m$_{AB} =$ 31.5, or
M$_{UV} =$ $-$15.5 (for $z =$ 7).  At $z =$ 6, our observations (shown in Figure 3)
already reveal a population of galaxies capable of sustaining
reionization.  At $z =$ 7, the observed population is not bright
enough to sustain reionization, thus if reionization completes at $z
>$ 6, fainter galaxies are necessary, which should be revealed by
JWST.  At $z =$ 8, even JWST cannot image deep enough to see the
hypothetical extremely faint population necessary to complete
reionization at $z =$ 8.}
\vspace{5mm}
\end{figure*}

However, at higher redshift, there
appears to be some tension between our observations and their results,
in that the \citet{kuhlen12} study prefers a 50\% reionized universe
at a higher redshift than appears possible with the observations,
though it is marginally consistent with the integrated luminosity
functions.  However, we stress that the results from \citet{kuhlen12}
were tuned to match the observed luminosity functions.
Any tension likely arises from the {\it WMAP} constraint on the Thomson
scattering optical depth due to electrons, which we have not accounted
for.  However, as the {\it WMAP} constraint
prefers a redshift of instantaneous reionization of 10.6 $\pm$ 1.2, it
is not surprising that observations showing a strong decline in the
galaxy UV luminosity density at $z \geq$ 8 would underproduce the electron optical
depth.  This can be reconciled by allowing
the escape fraction, and/or the limiting magnitude to evolve with
redshift, as is done by \citet{kuhlen12}.  
This would imply that very faint galaxies may be necessary to drive reionization at the earliest
stages, and that reionization gets its start at much higher redshifts
(i.e., $z >$ 10).  We point out that the $z =$ 8 luminosity function, and the limits on the
$z =$ 10 luminosity function, are very rough at this point in time.
Future observations, likely with {\it JWST} (see \S 4.7), will greatly
improve constraints on the luminosity functions in this early epoch,
and at that time we can learn more about whether the contribution of
galaxies to the early stages of reionization are consistent with all
other constraints.

\subsection{Implications for Future Surveys}

At $z =$ 6, our observations (shown in Figures 3 and 4) already reveal a population of galaxies capable of sustaining
reionization, while at $z =$ 7 and 8 the observed galaxy population
falls short.  Our current observations are however susceptible to the
uncertainty about what lies below our detection threshold, as well as
our assumptions about the escape fraction.  The former can potentially
be alleviated by integrating down the luminosity function, as we have
done in \S 4.4, but this yields its own uncertainty: how faint does
the UV luminosity function continue?

In order to measure this, we will require the deeper observations only
possible with {\it JWST}.  In a hypothetical $\sim$800 hr program,
JWST can reach m$_{AB} (5\sigma) =$ 31.5
mag in four bands; F070W, F090W, F115W and F150W; which can be used to
select samples at similar redshifts as in our
study\footnote[3]{http://jwstetc.stsci.edu/etc/input/nircam/imaging/}.
Is this deep enough to probe faint enough values of M$_{lim}$ to conclusively find
the reionizing population at $z =$ 7 and 8?  To investigate this, we
created Figure 7, where we plot the limiting absolute UV magnitude
necessary to sustain reionization for a given ratio of C/f$_{esc}$,
assuming the luminosity functions from the literature (see
\citet{wilkins11b} for a similar analysis at $z =$ 7).

In this figure, we show the limiting magnitude of our {\it HST} study,
as well as that of the hypothetical {\it JWST} deep field, where each
colored line denotes the value of M$_{lim}$ and C/f$_{esc}$ necessary
for a given redshift.  As we discussed above, for our assumption of
C/f$_{esc} =$ 10, our $z =$ 6 observed galaxy population is sufficient
to sustain a reionized IGM.  In fact, this
figure shows that the observed population can reionize the IGM for
values of C/f$_{esc} \lesssim$ 14.  At $z =$ 6, {\it JWST} will trace
the rest-frame UV luminosity function down to faint levels,
enabling the search for a turnover at faint magnitudes down to
M$_{1500} = -$15.3.  If the luminosity function continues its steep
climb to this magnitude, then as shown in Figure 5, the average escape
fraction for galaxies will be constrained to be less than $\sim$15\%.

At $z =$ 7, galaxies brighter than {\it HST}s magnitude limit can only
reionize the universe if C/f$_{esc} \lesssim$ 6, which, for C $\geq$
3, necessitates an escape fraction $\geq$ 50\%.  While this is
possible, there is not yet strong observational evidence to support
such a scenario.  However, {\it JWST} will reach deep enough, down to
M$_{lim} = -$15.5; if the luminosity function continues its steep
slope down to such faint magnitudes, and C/f$_{esc} \lesssim$ 15, then
galaxies could have reionized the IGM by $z =$ 7 \citep[see also][]{salvaterra11}.  At $z =$ 8, even {\it
  JWST} can not probe deep enough to see the limiting magnitude
necessary to reionize the IGM (again assuming C/f$_{esc} \geq$ 10).
If the universe was reionized by $z =$ 8, it necessitates a luminosity
function that continues with a steep slope ($\alpha \sim -$2) down to
M$_{lim} = -$12.  However, such a result would create tension with the
observed dearth of \lya\ emission, and the small size of quasar near
zones at $z =$ 7 (see \S 4.2).  This tension could be resolved if the
escape fraction of galaxies is luminosity dependent, and galaxies with
L $\ll$ $L^{\ast}$  have very low escape fractions, which could be the
case if they are highly gas dominated \citep[e.g.,][]{gnedin08}.

\section{Conclusions}

We have measured the specific UV luminosity density ($\rho_{UV}$) from a large
sample of galaxies at $z =$ 6, 7 and 8.  Using a theoretical model to
convert from UV to ionizing photons, and determining the level of
ionizing photons necessary to sustain a fully reionized IGM, we have
investigated the contribution of observed galaxies to the reionization
of the IGM.  A large body of work over the past $\sim$ decade points
to a probable low clumping factor.  Thus, previous surveys which assumed a large clumping factor and 
 deduced that galaxies could not reionize the universe may have
 reached the opposite conclusion today.  Unfortunately, not much is
 known about the escape fraction of ionizing photons at high
 redshift, but current observations are consistent with a rising
 escape fraction with increasing redshift, or at least an increased
 incidence of large escape fractions.  With our new sample, and our
 increased understanding of the process of reionization, we have reached the following conclusions:

\begin{itemize}
\item{We measure $\rho_{UV}$ at $z =$ 6 to be 0.79 $\pm$ 0.06 $\times$
    10$^{26}$ erg s$^{-1}$ Hz$^{-1}$ Mpc$^{-3}$ for the observed
    population of galaxies; this is
    much more robust than previous measures, which had much smaller
    samples and were limited to single band detections, and differed
    by a factor of $>$ 5.  Additionally, this result is independent of
    a Schechter function parametrization for the form of the
    luminosity function.  Assuming C/f$_{esc} \sim$ 10, we find that the observed galaxy population
    at $z =$ 6 is sufficient to sustain a fully reionized IGM.  In
    contrast to many previous studies, this result
    does not need to invoke galaxies fainter than our detection
    threshold.}

\item{We measure $\rho_{UV}$ of the observed galaxy population at $z =$ 7 and 8 to be 0.44 $\pm$ 0.04 and 0.27 $\pm$ 0.06 $\times$
    10$^{26}$ erg s$^{-1}$ Hz$^{-1}$ Mpc$^{-3}$, respectively.  These values fall short of
    what is needed to reionize the IGM.  While fainter galaxies may
    render the population capable of completing reionization, even
    {\it JWST} will not go deep enough to see the necessary limiting
    magnitude at $z =$ 8.}

\item{If we only consider the observed galaxies, we find a scenario
    where though $z =$ 7 and 8 galaxies contribute to reionization,
    the IGM only becomes fully reionized by $z =$ 6.  This is
    consistent with independent studies of both \lya\ emission from
    normal galaxies, and the study of \lya\ absorption in the
    near-zone of the $z =$ 7 quasar, both of which find that the IGM
    at $z =$ 7 likely has a neutral fraction $\geq$ 10\%.}

\item{Accounting for constraints on the total emission rate of
    ionizing photons from measurements of the \lya\ forest at $z =$ 6, we find
    that if our observed galaxies represent the total galaxy
    population, then the average $z =$ 6 escape fraction must be
    f$_{esc}$ $<$ 34\%.
    However, fainter galaxies likely do exist, and if the luminosity
    function extends down to M$_{lim} = -$13, then the average escape
    fraction constraint tightens to f$_{esc}$ $<$ 13\%.  These escape fractions
    can sustain reionization $z =$ 6 either with our observed galaxies, or
    with the integrated luminosity function, for values of the
    clumping factor $\leq$ 3.}

\item{We investigate the evolution of the volume ionized fraction
    x$_{HII}$ of the IGM.  We do this by combining the observed galaxy population of
  our study and the luminosity functions from the literature with
  limits on the escape fraction considering both limiting magnitudes.
While the IGM appears to be fully ionized by $z \leq$ 6, the volume
ionized fraction appears to drop below unity by $z =$ 7, consistent
with a number of complementary analyses.  Considering only the
(roughly known) contribution from galaxies at $z \geq$ 8, the volume
ionized fraction may drop substantially below unity, though this is in
mild tension with reionization models which incorporate the optical
depth due to electron scattering.  Better constraints on the $z \geq$
8 luminosity functions, likely from {\it JWST}, will place stronger
constraints on the reionization history at that early time.  If the
strong drop in the luminosity density at $z >$ 8 is verified, it may
imply some additional source of ionizing photons at $z >$ 10 is
warranted to remain consistent with the {\it WMAP}-measured Thomson scattering
optical depth.}

\end{itemize}

We conclude that in order to make progress on the issue of when and
how reionization happened, we need better constraints on the escape
fraction of ionizing photons or the limiting magnitude of the UV
luminosity function.  As these quantities are correlated due to the
constraints from the \lya\ forest, constraints on one will yield more
robust constraints on the other.  While direct measurement of the
escape fraction at high redshift is not possible, much deeper
observations with {\it JWST} should confirm whether the luminosity
function extends down to at least M$_{lim} = -$15.5, which will place
stronger constraints on the escape fraction at $z =$ 6, and uncover
whether the $z =$ 7 galaxy population is sufficient to reionize the universe.

\acknowledgements
\vspace{-2mm}
We thank Volker Bromm, Andrea Ferrara, and Milos Milosavljevic for
stimulating conversations, as well as Chris Conselice, Yan Gong,
Bahram Mobasher and Brian Siana for useful comments.  Support for SLF
was provided by NASA through Hubble Fellowship grant HST-HF-51288.01.
CP was supported in part by {\it HST} program 12060.  These grants
were all awarded by the Space Telescope Science Institute, which is
operated by the Association of Universities for Research in Astronomy,
Inc., for NASA, under contract NAS 5-26555.  JSD acknowledges the
support of the European Research Council through an Advanced Grant, 
and the support of the Royal Society via a Wolfson Research Merit award.

\end{document}